\documentclass[a4paper,reprint,twocolumn,aps,pra, superscriptaddress,floatfix]{revtex4-2}
\usepackage{graphicx,booktabs,multirow,longtable,upgreek,hyperref,amsmath,bm} 
\usepackage[shortlabels]{enumitem}
\usepackage[dvipsnames]{xcolor}
\usepackage{tikz}
\usetikzlibrary{arrows.meta}
\usetikzlibrary{decorations.pathreplacing}
\usetikzlibrary{shapes.geometric}

\newcommand{\rme}{\mathrm{e}}

\newcommand{\rmd}{\mathrm{d}}
\newcommand{\rr}{\bm{r}}

\newcommand{\Ecum}{E_{\mathrm{cum}}}
\newcommand{\Sph}{S_\mathrm{ph}}
\newcommand{\Exc}{E_{\mathrm{xc}}}
\newcommand{\EH}{E_{\mathrm{H}}}
\newcommand{\Exchyb}{E_{\mathrm{xc}}^\mathrm{hyb}}

\newcommand{\ExDFA}{E_{\mathrm{x}}^\mathrm{DFA}}
\newcommand{\EcDFA}{E_{\mathrm{c}}^\mathrm{DFA}}
\newcommand{\ExHF}{E_{\mathrm{x}}^\mathrm{HF}}
\newcommand{\Exchybtheta}{E_{\mathrm{xc}}^{\mathrm{hyb}+\theta}}
\newcommand{\ExHFtheta}{E_{\mathrm{x}}^{\mathrm{HF}+\theta}}
\newcommand{\vxDFA}{v_{\mathrm{x}}^\mathrm{DFA}}
\newcommand{\vcDFA}{v_{\mathrm{c}}^\mathrm{DFA}}
\newcommand{\vext}{v_\mathrm{ext}}
\newcommand{\veff}{v_\mathrm{eff}}
\DeclareMathOperator{\trace}{Tr}
\DeclareMathOperator*{\argmin}{argmin}
\newcommand{\corrtemp}{+\theta}
\newcommand{\bra}[1]{\mbox{$\langle #1 |$}}
\newcommand{\ket}[1]{\mbox{$| #1 \rangle$}}

\DeclareRobustCommand{\markerstar}{%
  \tikz[baseline={([yshift=-2.7pt]char.center)}]{%
    \node[star, star points=5, star point ratio=2.5, draw, inner sep=0pt, minimum size=10pt, fill=white] (char) {};
  }%
}
\DeclareRobustCommand{\markerdiamond}{%
  \tikz[baseline={([yshift=-2.7pt]char.center)}]{%
    \node[draw, diamond, inner sep=1pt, minimum size=8pt, fill=white] (char) {};
  }%
}
\DeclareRobustCommand{\markersquare}{%
  \tikz[baseline={([yshift=-2.7pt]char.center)}]{%
    \node[draw, rectangle, inner sep=1pt, minimum size=6pt, fill=white] (char) {};
  }%
}

\begin{document}

\author{Florian Matz}
\affiliation{Arnold Sommerfeld Centre for Theoretical Physics, Ludwig-Maximilians-Universität München, Munich, Germany}

\author{Paul G. Graf}
\affiliation{Arnold Sommerfeld Centre for Theoretical Physics, Ludwig-Maximilians-Universität München, Munich, Germany}

\author{Markus Penz}
\affiliation{Arnold Sommerfeld Centre for Theoretical Physics, Ludwig-Maximilians-Universität München, Munich, Germany}
\affiliation{Department of Computer Science, Oslo 
Metropolitan University, Oslo, Norway}

\author{Christian Schilling}
\email{christian.schilling@unige.ch}
\affiliation{Department of Physical Chemistry, University of Geneva, Geneva, Switzerland}
\affiliation{Arnold Sommerfeld Centre for Theoretical Physics, Ludwig-Maximilians-Universität München, Munich, Germany}

\date{\today}

\title{Adding Temperature to the Adiabatic Connection: Hybrid Functionals for Static and Dynamic Correlation}

\begin{abstract}
    Capturing static and dynamic electron correlation simultaneously at manageable computational cost remains a key challenge in electronic structure theory. Motivated by the recurring role of effective temperature in self-consistent field methods we extend the adiabatic connection mechanism from density functional theory (DFT) to include a temperature dimension. The formalism unifies established hybrid functionals with information-density matrix functional theory (i-DMFT) and thermally-assisted-occupation DFT, clarifying their theoretical basis and the respective prescriptions for the temperature parameter. It further gives rise to thermally enhanced hybrid functionals, in which static correlation is captured via fractional occupations generated by an effective ``correlation temperature'' within thermal Hartree-Fock, while dynamic correlation is described by established density functional approximations. We assess the performance of several thermally enhanced hybrids for different cases where i-DMFT was previously shown to fail. Furthermore, we discuss and test different schemes for determining the correlation temperature and investigate the accuracy of different such choices for a benchmark set of different organic molecules. The results clarify the achievable balance of accuracy, generality, and efficiency, establishing thermally enhanced hybrids as a systematic route toward low-cost treatment of both correlation types.
\end{abstract}

\maketitle

\section{Introduction}
\label{sec:intro}

The toolbox of computational chemistry is vast and has expanded significantly over the last decades. The determining challenge today is to find efficient and simple-to-use methods for large molecules and large datasets, while still taking into account the complexity of different electronic structures, depending significantly on molecular composition. 
The accurate description of molecular electronic structure requires methods that can handle both static and dynamic electron correlation.~\cite{borden96,Cremer2001,benavidesriveros17,vianadal19} Static correlation arises from near-degenerate configurations and requires a multireference treatment~\cite{keller15,knecht16,coe15,AlivertiPiuri2024,zhou22}; dynamic correlation captures the remaining electron-electron interactions~\cite{borden96,stein16} and is often treatable perturbatively or via density functionals. While coupled-cluster~\cite{shavitt2009many,bartlett07,bartlett11} and perturbation theories~\cite{moller34,cremer11} handle dynamic correlation efficiently for single-reference systems, and DFT offers a practical mean-field framework~\cite{hohenberg64,yu16,verma20,Burke2012}, no established method simultaneously handles both types of correlation accurately, efficiently, and systematically.~\cite{andersson92,zhou22,koehn12,kinoshita05,krylov08}

Different correlation mechanisms are not only a phenomenological effect in quantum-chemical calculations. Instead, understanding and describing correlation in systems of interacting fermions are also possible through the plethora of tools quantum information theory has to offer. Static correlation corresponds to entanglement of electrons or the orbitals with which quantum chemistry describes them~\cite{Gottlieb2005,ding20,Graf2026}, while dynamic correlation can be connected to the tail of small occupation numbers that never exactly vanish.~\cite{benavidesriveros17,giesbertz2018, Giesbertz2013}

An attractive framework to express fermionic wave functions without relying on the explicit construction of a large amount of configurations is reduced density-matrix functional theory (RDMFT).~\cite{davidson-book,Gilbert75,Valone80,M84,Schilling18,Gibney2022,liebert23,Fredheim2025} Therein, a state $|\Psi\rangle$ is reduced to its one-particle density matrix (1RDM)
\begin{equation}
    \gamma_{ij} = \langle\Psi|\hat a_i^\dagger\hat a_j|\Psi\rangle.    
\end{equation}
Just like with the one-particle density, a mapping of the 1RDM to the exact electronic ground-state energy is in principle possible. While the exact functional that achieves this is not known and variation over all possible 1RDMs can come with intricacies, RDMFT already inspires quantum-chemical approaches:
Fractional natural occupation numbers, as the eigenvalues of the 1RDM, offer a promising route to capturing static correlation at mean-field cost. The i-DMFT method~\cite{WangBaerends22-PRL} achieves this via a conjectured linear relation between cumulant energy and particle-hole symmetrized entropy~\cite{Collins,Ziesche1995,WangKnowles21-PRA}, leading to a thermal Hartree-Fock (HF) picture.~\cite{Mermin1963}
Thermally-assisted-occupation DFT (TAO-DFT)~\cite{Chai2012,tsai13,Chai2014,Chai2017} pursues a similar idea within DFT, augmenting standard functionals with Fermi-Dirac occupations governed by an electronic temperature $\theta$. In both cases, the central challenge is determining $\theta$ appropriately.
Various approaches for this have been proposed, from matching high-level occupation numbers~\cite{Chen2022} to orbital-energy based criteria.~\cite{Lin2017} These findings also inspire recent efforts: \Citet{Feng2025} proposed a combination of (pair\nobreakdash-)DFT and entropy-based correlation functionals, where the total electronic energy is evaluated from the density, pair density, and occupation numbers obtained from a CASSCF reference calculation.
We recently showed that there is an intimate connection between correlation effects, correlation measures, and correlation temperature.~\cite{perspective} However, a recent assessment~\cite{Graf2026} revealed persistent failures of i-DMFT for dissociation barriers especially in case of changing electronic character, motivating a more rigorous theoretical foundation and numerical testing.

This repeated occurrence of entropy and effective temperature in electronic structure methods motivated us to investigate the theoretical foundations of those methods, in particular of DFT. It turns out that DFT itself can be formulated as a \emph{thermodynamic} theory. This viewpoint (Section~\ref{sec:dft-thdyn}) then leads naturally to a novel adiabatic connection that extends the usual coupling-strength integration~\cite{Gunnarsson1976,Becke1993a,Becke1993b,PEB1996,Adamo1999,Toulouse2022} by a temperature dimension, providing a rigorous explanation of why an electronic temperature and exact HF exchange emerge simultaneously. The resulting \emph{thermally enhanced} hybrid functionals (Section~\ref{sec:thermal-hybrid}) cover static and dynamic correlation and the transition between them, while operating at mean-field cost. This allows to treat the relatively simple features of static correlation by introducing an effective temperature, while dynamical correlation is still represented in the total electron density.
By systematically translating this into a numerical method (Section~\ref{sec:num-method}), we demonstrate how known failures of i-DMFT in densities, occupation numbers, and dissociation barriers are substantially reduced (Section~\ref{sec:study-dissoc}). We critically discuss existing strategies for determining correlation temperatures, arguing that neither fully universal values nor highly system-specific methodologies are satisfactory, and we propose alternatives (Section~\ref{sec:theta-approaches}). Benchmarking on a set of organic molecules then tests the transferability of these ideas across chemically similar systems (Section~\ref{sec:larger}). We conclude with an outlook on future development and applications (Section~\ref{sec:conclusions}).

\section{DFT as thermodynamics}
\label{sec:dft-thdyn}

DFT, as a quantum chemistry method used to investigate the electronic structure of many-body systems in their equilibrium state, operates by switching from the wavefunction to the electron density as a reduced variable. Its mathematical structure is not merely reminiscent of thermodynamics, it in a sense \emph{is} thermodynamics.~\cite{Argaman2000,Argaman2002} We demonstrate this claim here by deriving DFT in a thermodynamic way. This will then form the theoretical basis for our proposed thermally enhanced hybrid approximation.

Naturally, for this purpose, we study systems with positive temperature, $\tau>0$, but the usual zero-temperature case is also included as a well-defined limit. The system will be described by the Hamiltonian
\begin{equation}
    \hat H^{\lambda}[v]=\hat T+\lambda\hat W+\hat V[v],
\end{equation}
where $\hat T$ is the kinetic part (that could include other fixed one-body parts, like an overall containment potential), $\hat W$ is the Coulomb interaction with interaction strength $\lambda\geq 0$, and the external potential $\hat V$ is given by the one-particle potential $v$. This means $(\lambda,v)$ are a parametrization of a family of Hamiltonians. Adding temperature to this and keeping the particle number variable means that we do not minimize the expectation value of the Hamiltonian, but instead use the trial grand potential~\cite{mermin1965thermal}
\begin{equation}\label{eq:grand-pot}
    \Omega^{\lambda,\tau}(v,\hat\Gamma) = \trace\left(\hat\Gamma \hat H^{\lambda}[v]\right) + \tau\trace\left(\hat\Gamma\log\hat\Gamma\right)
\end{equation}
that now has parametrization $(\lambda,\tau,v)$ and gets minimized over all ensemble states $\hat\Gamma$ from the Fock space. The second term equals minus the von Neumann entropy times the temperature. In this work, we will generally consider grand canonical ensembles, which means the chemical potential $\mu$ actually enters as an additional parameter, but it can always be included in the external potential $v$ as a constant shift and will further not be denoted explicitly.
At $\tau=0$ this recovers the usual energy of the system, $E^\lambda=\Omega^{\lambda,0}$, and the system can then be in a pure ground state with zero particle number fluctuation. 
For $\tau>0$, the minimizer of Eq.~\eqref{eq:grand-pot} is always uniquely given by the Gibbs ensemble
\begin{equation}
    \hat\Gamma^{\lambda,\tau}[v] = \frac{\rme^{-\hat H^\lambda[v]/\tau}}{Z},\quad Z=\trace\left( \rme^{-\hat H^\lambda[v]/\tau} \right).
\end{equation}

The attained minimum of the grand potential in Eq.~\eqref{eq:grand-pot} for the given set of parameters then defines the functional $\Omega^{\lambda,\tau}(v)$ that assumes the role of a thermodynamic potential. It is found to be concave in $v$ and Legendre transform then allows one to switch the potential to its dual variable, the one-particle density $\rho$ and define a convex density functional as the free energy~\cite{eschrig2010},
\begin{equation}\label{eq:dens-func}
    F^{\lambda,\tau}(\rho) = \sup_v \left\{ \Omega^{\lambda,\tau}(v) - \int v(\rr)\rho(\rr)\rmd\rr \right\}.
\end{equation}
The density $\rho$ and its dual variable $v$ are naturally position-dependent, but could also be taken spin-dependent with an additional spin variable $\sigma\in\{\uparrow,\downarrow\}$, thus extending the formalism to spin-DFT~\cite{Barth1972}, without changing anything in the formalism.
The important zero temperature case is included in the whole discussion as the limit $\tau\to 0$~\cite[Eq.~(10)]{ullrich2002}, where Eq.~\eqref{eq:dens-func} becomes the ensemble functional of usual DFT.~\cite{Lieb1983}

In order to relate the functional $F^{\lambda,\tau}(\rho)$ at some density $\rho$ to a reference at different parameter values $(\lambda,\tau)$, we take the density $\rho$ as fixed and further make the critical assumption that for all parameter values $(\lambda,\tau)$ of interest there is an external potential $v^{\lambda,\tau}$ such that $\rho$ is the corresponding equilibrium density (``$v$-representability''). This is by no means a trivial assumption, and it has only been proven to be guaranteed in very limited settings (\citet{Sutter2026} give a solution for the canonical ensemble on a one-dimensional torus, while the $\tau=0$ case was treated before elsewhere~\cite{sutter2024solution,Corso2025,Corso2026a}).
Note that this assumption, through the existence of a Gibbs ensemble, also implies that the partition function $Z$ is finite, which is not assured at all for chemical systems and even fails for the hydrogen atom.~\cite[Section~2.3]{GiesbertzRuggenthaler2019} Here, having the density $\rho$ fixed works as if the volume of the system would be bounded, which is the typical requirement of thermodynamics and a necessary condition for a finite partition function.
In the following, we will omit this fixed $\rho$ from the notation as an explicit argument.
The assumption of $v$-representability allows one to find a corresponding ensemble $\hat\Gamma^{\lambda,\tau}[v^{\lambda,\tau}]$ that has density $\rho$ and allows one to define the individual energies
\begin{align}
    \label{eq:T-def}
    T^{\lambda,\tau} &= \trace\left( \hat\Gamma^{\lambda,\tau}[v^{\lambda,\tau}]\hat T \right)\\
    \label{eq:W-def}
    W^{\lambda,\tau} &= \trace\left( \hat\Gamma^{\lambda,\tau}[v^{\lambda,\tau}]\hat W \right)
\end{align}
and the entropy
\begin{equation}
    \label{eq:S-def}
    S^{\lambda,\tau} = -\trace\left( \hat\Gamma^{\lambda,\tau}[v^{\lambda,\tau}]\log \hat\Gamma^{\lambda,\tau}[v^{\lambda,\tau}] \right).
\end{equation}

For $\tau>0$ the ensemble $\hat\Gamma^{\lambda,\tau}[v^{\lambda,\tau}]$ is unique~\cite{eschrig2010}, but even at $\tau=0$ the above quantities from Eqs.~\eqref{eq:T-def} and \eqref{eq:W-def} are uniquely given by virtue of the virial relation for chemical systems.~\cite[Section~II]{Laestadius2024-VRx} This now allows one to write
\begin{equation}\label{eq:functional-parts}
    F^{\lambda,\tau} = T^{\lambda,\tau} + \lambda W^{\lambda,\tau} - \tau S^{\lambda,\tau}
\end{equation}
and to retrieve the grand potential by Legendre back-transformation
\begin{equation}
    \Omega^{\lambda,\tau}(v) = \inf_\rho \left\{ F^{\lambda,\tau}(\rho) + \int v(\rr)\rho(\rr)\rmd\rr \right\}.
\end{equation}
As seen from the equation above, the representing potential $v^{\lambda,\tau}$ for the fixed density can be retrieved by functional differentiation of the density functional,
\begin{equation}
    -v^{\lambda,\tau} = \frac{\delta F^{\lambda,\tau}}{\delta \rho}.
\end{equation}
After assuming $v$-representability, this potential is always unique if $\tau>0$~\cite{eschrig2010}, while at $\tau=0$ uniqueness is the content of the Hohenberg-Kohn theorem~\cite{hohenberg64} that remains a delicate question despite substantial recent progress.~\cite{Garrigue2018,Corso2026b}

The principle of duality for thermodynamic quantities extends also to the other parameters, which can be seen as follows. For different $\lambda,\lambda'$ it follows from the variational principle of the grand potential (\citet[Section~III]{Laestadius2024-VRx} have this analogously for the usual density functional) that
\begin{equation}\begin{aligned}
    F^{\lambda',\tau} &= T^{\lambda',\tau} + \lambda' W^{\lambda',\tau} - \tau S^{\lambda',\tau} \\
    &\leq T^{\lambda,\tau} + \lambda' W^{\lambda,\tau} - \tau S^{\lambda,\tau} \\
    &= F^{\lambda,\tau} + (\lambda'-\lambda)W^{\lambda,\tau}.
\end{aligned}\end{equation}
By definition, this means that $W^{\lambda,\tau}$ is in the superdifferential of $F^{\lambda,\tau}$ with respect to $\lambda$.
Swapping $\lambda$ and $\lambda'$ gives $F^{\lambda,\tau}\leq F^{\lambda',\tau} + (\lambda-\lambda')W^{\lambda',\tau}$ and addition of the two inequalities yields
\begin{equation}
    0\leq (\lambda'-\lambda)(W^{\lambda,\tau}-W^{\lambda',\tau}).
\end{equation}
So $\lambda'>\lambda$ implies $W^{\lambda,\tau}\geq W^{\lambda',\tau}$, which means $W^{\lambda,\tau}$ is nonincreasing in $\lambda$.~\cite{Levy1985} (The conjectured convexity of $W^{\lambda,0}$ is still a major open problem in DFT.~\cite{Crisostomo2023})
Analogously it follows that $-S^{\lambda,\tau}$ is in the superdifferential of $F^{\lambda,\tau}$ with respect to $\tau$ and that $S^{\lambda,\tau}$ is nondecreasing in $\tau$. The derivative of $S^{\lambda,\tau}$ in $\tau$ relates to the (nonnegative) heat capacity of the system. Written with a usual partial derivative, this establishes the thermodynamic relations
\begin{equation}\label{eq:A-derivatives}
    W^{\lambda,\tau} = \frac{\partial F^{\lambda,\tau}}{\partial \lambda} \quad\text{and}\quad -S^{\lambda,\tau} = \frac{\partial F^{\lambda,\tau}}{\partial \tau}.
\end{equation}

From the fundamental theorem of integral calculus (that equally applies to superdifferentials~\cite{niculescu-persson-book}) it now follows for \emph{any} path $C$ in $(\lambda,\tau)$ parameter space that connects points $(\lambda_1,\tau_1)$ and $(\lambda_2,\tau_2)$ that
\begin{equation}\label{eq:AC}
    F^{\lambda_2,\tau_2} = F^{\lambda_1,\tau_1} + \int_C \left(W^{\lambda,\tau} \rmd \lambda - S^{\lambda,\tau} \rmd \tau\right).
\end{equation}
This defines a \emph{thermally enhanced adiabatic connection} between two systems at $(\lambda_1,\tau_1)$ and $(\lambda_2,\tau_2)$, where the usual adiabatic connection of DFT is between the Kohn-Sham (KS) system at $(0,0)$ and the fully interacting system $(1,0)$.
Very recently, \citet{han2026} formulated the similar idea of a ``finite-temperature adiabatic connection'', which also aims at placing i-DMFT and TAO-DFT into a common framework but establishes the connection at fixed 1RDM rather than at fixed density.
Note that the word `adiabatic' does not have the same meaning here as in thermodynamics, where it describes state changes without heat exchange, but instead refers to changes that are slow enough such that the system remains in an  instantaneous equilibrium and can consequently always be described by the Gibbs state.

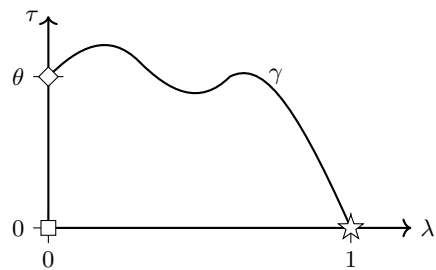
\begin{figure}[ht]
\centering
\begin{tikzpicture}[scale=4]
    \draw[->, thick] (0,0) -- (1.2,0) node[right] {$\lambda$}; \draw[->, thick] (0,0) -- (0,0.7) node[left] {$\tau$};     \foreach \x in {0,1} {
        \draw (\x,0.05) -- (\x,-0.05);
    }
    \node[below] at (0,-0.05) {$0$};
    \node[below] at (1,-0.05) {$1$};

    \foreach \y in {0,0.5} {
        \draw (0.05,\y) -- (-0.05,\y);
    }
    \node[left] at (-0.05,0) {$0$};
    \node[left] at (-0.05,0.5) {$\theta$};

    \draw[thick, smooth, looseness=1.0] 
        (0,0.5) 
        .. controls (0.1,0.6) and (0.2,0.65) .. 
        (0.3,0.55) 
        .. controls (0.4,0.45) and (0.5,0.4) .. 
        (0.6,0.5)
        .. controls (0.7,0.55) and (0.8,0.45) .. 
        (1,0);
    \node[right] at (0.75,0.5) {$C$};

    \node[star, star points=5, star point ratio=2.5, draw, scale=0.45, fill=white] at (1,0) {};
    \node[rectangle, draw, scale=0.9, fill=white] at (0,0) {};
    \node[diamond, draw, scale=0.6, fill=white] at (0,0.5) {};
\end{tikzpicture}
\caption{A general path $C$ in $(\lambda,\tau)$ parameter space that connects a thermal KS system (\markerdiamond) at $(0,\theta)$ with the fully interacting system (\markerstar) at $(1,0)$ while keeping the density fixed. The usual KS system (\markersquare) is located at the origin $(0,0)$.}
\label{fig:path}
\end{figure}

The idea for (thermally enhanced) functional approximations is now the following. We can choose any reference system in the parameter space $(\lambda,\tau)$ that seems suitable, in practice this will of course mean setting the interaction strength $\lambda=0$, and then exactly follow the adiabatic connection to the interacting system of interest. The corresponding Hartree-exchange-correlation (Hxc) functional is then given precisely by the integral in Eq.~\eqref{eq:AC} and can be further approximated from there. But as demonstrated elsewhere~\cite{perspective,ramos2019}, an effective temperature in a non-interacting system alone can already serve as a viable approximation to correlation effects, so it makes sense to not take the usual KS system at $(\lambda,\tau)=(0,0)$ as a reference, but instead move to thermal KS at $(0,\theta>0)$ and move to the interacting system from there along an adiabatic connection path. The corresponding diagram with an arbitrary path is illustrated in Fig.~\ref{fig:path}, but of course, we will in practice follow a path that is much easier to control. This will be elaborated in Section~\ref{sec:thermal-hybrid}.
Independent of the chosen path $C$ between $(0,\theta)$ and $(1,0)$ we can then define the exchange-correlation (xc) energy functional from Eq.~\eqref{eq:AC} as
\begin{equation}\begin{split}
    \Exc^\theta &= F^{1,0} - F^{0,\theta} - \EH \\
    &= \int_C \left(W^{\lambda,\tau} \rmd \lambda - S^{\lambda,\tau} \rmd \tau\right) - \EH.
\end{split}
\end{equation}
We can now follow this path in two steps, first from $(0,\theta)$ to the KS point $(0,0)$, and then further to the interacting system at $(1,0)$, to find
\begin{equation}\label{eq:Exc-S-int}
\begin{aligned}
    \Exc^\theta &= \int_0^\theta S^{0,\tau} \rmd\tau + \int_0^1 W^{\lambda,0} \rmd\lambda - \EH \\
    &= \Exc + \underbrace{ \int_0^\theta S^{0,\tau} \rmd\tau }_{\textstyle E_\theta}.
\end{aligned}
\end{equation}
Here, we used that the usual adiabatic connection integral over the interaction strength yields $F^{1,0}-F^{0,0}=\EH+\Exc$ and thus recovers the usual xc functional from DFT. This allows to precisely connect the new thermally enhanced $\Exc^\theta$ and $\Exc$ with an additive term that is exactly the $E_\theta$ functional from TAO-DFT~\cite{Chai2012}, and we notice that it is defined as an adiabatic-connection integration of entropy over temperature.

Note that the relations closely resemble those in finite-temperature DFT~\cite{Pittalis2011,burke2016thDFT}, with the critical difference that the interacting system and the KS system remain at the same temperature there. A thermal connection formula that involves coordinate scaling instead appeared previously in \citet{PribramJones2016}. A connection formula that integrates the full rectangle between $(0,\theta)$ and $(1,0)$ instead of the path connecting the points was considered in \citet{Harding2024}.

\section{Thermally enhanced hybrid functionals}
\label{sec:thermal-hybrid}

The basic idea behind hybrid functionals~\cite{Becke1993a,Becke1993b,PEB1996} is that common density functional approximations (DFA) are more accurate close to $\lambda=1$, while at $\lambda=0$ one remains in the domain of exact exchange in first order of $\lambda$ (this is a precise result visible from Eq.~\eqref{eq:A-derivatives} by just choosing $(\lambda,\tau)=(0,0)$). The rationale is then to mix in a fraction of exact (HF-like) exchange with DFA exchange functionals, while the correlation part of the DFA stays untouched. This helps to achieve more accurate results while staying entirely within the DFT framework thanks to the adiabatic connection.

We first review the zero-temperature case of standard hybrid functionals and start from the exact relation
\begin{equation}
    \Exc = \int_0^1 W^{\lambda,0}\rmd\lambda - \EH
\end{equation}
already used in Eq.~\eqref{eq:Exc-S-int}.
A simple approximation to this integral would just interpolate the exchange contribution between the integrand's value at $\lambda=0$ (HF exact exchange) and $\lambda=1$ (approximated via DFA). This motivates the standard form of hybrid functionals in DFT,
\begin{equation}
    \Exc\approx\Exchyb = a\ExHF + (1-a)\ExDFA + \EcDFA,
\end{equation}
where the value for $a$ ranges from $0.20$~\cite{Becke1993b} to $0.25$~\cite{PEB1996,Adamo1999} and further to higher values for specialized functionals. For an overview of popular hybrid functionals of the above type we refer to \citet[Section~4.1]{Toulouse2022}. In the following we will build on this well established family of DFAs for our construction.

\begin{figure}[ht]
\centering
\begin{tikzpicture}[scale=5.5]
    \draw[->, thick] (0,0) -- (1.2,0) node[right] {$\lambda$}; \draw[->, thick] (0,0) -- (0,0.7) node[left] {$\tau$};
    \foreach \x in {0,0.25,1} {
        \draw (\x,0.05) -- (\x,-0.05);
    }
    \node[below] at (0,-0.05) {$0$};
    \node[below] at (1,-0.05) {$1$};
    \draw[dashed] (0.25,0.05) -- (0.25,0.5);
    \foreach \y in {0,0.5} {
        \draw (0.05,\y) -- (-0.05,\y);
    }
    \node[left] at (-0.05,0) {$0$};
    \node[left] at (-0.05,0.5) {$\theta$};
    \draw[blue, thick] (0,0.5) -- (0.25,0.5);
    \draw[red, thick] (0.25,0.5) -- (1,0);
    \draw[blue, thick] (0,0) -- (0.25,0);
    \draw[red, thick] (0.25,0) -- (1,0);
    \draw[blue, ->] (0.05,0.55) -- node[above, midway, font=\footnotesize] {$\frac{\partial}{\partial\lambda}\!F^{\lambda,\tau}$} (0.2,0.55);
    \draw[red, ->] (0.56,0.35) -- node[above, midway, sloped, font=\footnotesize] {$\frac{1}{L}\!\!\left[ (1-a)\frac{\partial}{\partial\lambda}-\theta\frac{\partial}{\partial\tau} \right]\!\!F^{\lambda,\tau}$} (0.71,0.25);
    \draw[decorate, decoration={brace, amplitude=5pt, mirror, raise=2pt}] (0,0) -- (0.25,0) node[midway, below=6pt] {$a$};
    \draw[decorate, decoration={brace, amplitude=5pt, mirror, raise=2pt}] (0.25,0) -- (1,0) node[midway, below=6pt] {$1-a$};
    \node[star, star points=5, star point ratio=2.5, draw, scale=0.45, fill=white] at (1,0) {};
    \node[rectangle, draw, scale=0.9, fill=white] at (0,0) {};
    \node[diamond, draw, scale=0.6, fill=white] at (0,0.5) {};
\end{tikzpicture}
\caption{The thermally enhanced hybrid is derived from the adiabatic connection linking a thermal KS system (\markerdiamond) with the fully interacting system (\markerstar). The usual hybrid starting at the KS system (\markersquare) is displayed along the $\lambda$-axis. The integrands for the two parts of the path are given in blue and red, with $L=\sqrt{(1-a)^2+\theta^2}$. For the whole blue part the integrand in the thermally enhanced adiabatic connection is approximated by the system at \markerdiamond{} (at \markersquare{} for usual hybrids), for the red part at \markerstar.}
\label{fig:path-hybrid}
\end{figure}
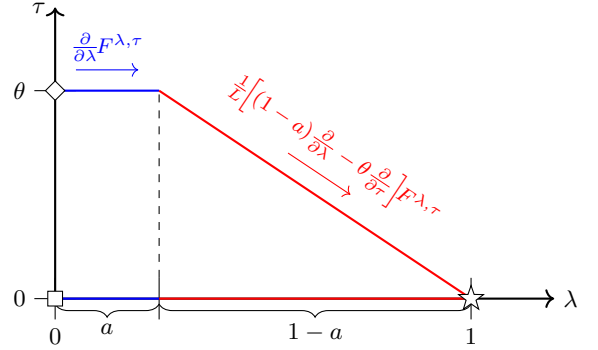

Exactly the same argument now applies to the ther\-mal\-ly-enhanced adiabatic connection, just that the path now leads from $(0,\theta)$ instead of $(0,0)$ to $(1,0)$. We choose the path displayed in Fig.~\ref{fig:path-hybrid} that starts off horizontally and then follows a straight line towards the interacting system.
The first (blue) part approximates the integrand fixed as $\frac{\partial}{\partial\lambda}F^{\lambda,\tau}=W^{\lambda,\tau}$ evaluated at $(0,\theta)$. Thus, integration over length $a$ adds $a\ExHFtheta$ (thermal HF exact exchange, as in \citet[Eq.~(2.5)]{Mermin1963}) to the functional. The second (red) part approximates the integrand as $\frac{1}{L}\left[ (1-a)\frac{\partial}{\partial\lambda}-\theta\frac{\partial}{\partial\tau} \right] F^{\lambda,\tau} = \frac{1}{L}\left[ (1-a)W^{\lambda,\tau}+\theta S^{\lambda,\tau} \right]$ evaluated at $(1,0)$ and integration over length $L$ yields $(1-a)\ExDFA$, just like in the usual hybrid functional, plus an entropy term.
The resulting approximation is
\begin{equation}\begin{aligned}\label{eq:dens-func-interacting}
    &F^{1,0}=F^{0,\theta}+\EH+\Exc^\theta \\
    &\! \approx F^{0,\theta} + \EH +a\ExHFtheta+(1-a)\ExDFA + \theta S^{1,0} + \EcDFA.
\end{aligned}\end{equation}
We note that the entropy term $S^{1,0}$ drops if \emph{pure-state interacting} $v$-representability can be assumed: in the case of $\tau=0$, the ground state is a rank-1 density matrix, which evaluates the von Neumann entropy to 0.
This results in a \emph{thermally enhanced hybrid functional} that mixes thermal HF with a DFA,
\begin{equation}\label{eq:Exctheta}
    \Exc^\theta \approx \Exchybtheta = a\ExHFtheta + (1-a)\ExDFA + \EcDFA.
\end{equation}
Although an exact integration along the path means that the respective choice of correlation temperature $\theta$ has no effect, it does matter within the approximation and enters the functional as an additional parameter.
The optimal temperature thus corresponds to the auxiliary reference that best captures the exact density functional or key physical properties via the approximate functional.

An immediate conceptual result is that this approximation neatly differentiates between static and dynamic correlation. Modern DFAs can handle dynamic correlation effects from short-range electron-electron repulsion relatively well using the local electron density. Conversely, they usually fail for strong static correlation, when several electronic configurations arise with almost equal weights.~\cite{Cremer2001,Cohen2008} So the load to handle the dynamic part of correlation is ceded to a DFT functional, while the first term in Eq.~\eqref{eq:Exctheta}, $a\ExHFtheta$, takes over the responsibility for static correlations with an adequate choice of effective temperature. One immediately realizes that the choices of parameters $a$ and $\theta$ are interlinked here, which also gets discussed in \citet{Chen2022}.
In the graphical depiction in Fig.~\ref{fig:path-hybrid} the parameters $a$ and $\theta$ determine the path for the adiabatic connection integral that gets approximated. In the following we discuss possible choices and how they reproduce known methods.

In the boundary case $a=0$ and setting the effective temperature $\theta=0$, the functional from \eqref{eq:Exctheta} is just a usual DFT functional.
With the particular choice $a=1$ and $\EcDFA=0$, on the other hand, the method retrieves the thermal HF method and approaches standard HF in the limit $\theta\to 0$ (albeit with a mix of orbitals at the Fermi level in case of degeneracy).
With the same choice $a=1$ and a constant $\EcDFA=-b$, the i-DMFT method~\cite{WangBaerends22-PRL} is exactly reproduced. The respective choice of parameters $\theta$ and $b$ then depends on the chemical scenario, like the atoms taking part in a dissociation process, but in certain cases was found to yield a good approximation over a variety of geometries.~\cite{Graf2026,WangBaerends22-PRL,Irimia2023-correlation}
With values $0<a<1$ one then truly enters the domain of thermally enhanced hybrid functionals. The limit $\theta\to 0$ means passing to a usual hybrid functional but again choosing a weighted average of densities in case of degeneracy. Table~\ref{tab:param-overview} gives an overview of possible parameter choices.

\begin{table}[htbp]
\centering
\caption{Overview of parameter choices in $\Exchybtheta$ (Eq.~\eqref{eq:Exctheta}) and the known methods this retrieves.}
\label{tab:param-overview}
\begin{tabular}{c|r|c|l}
\toprule
$a$ & $\theta$ & $\EcDFA$ & method \\
\midrule
$0$            & $0$  & standard DFA   & Kohn-Sham DFT \\
$1$            & $0$  & $0$            & Hartree-Fock \\
$1$            & $>0$ & $0$            & Thermal HF \\
$1$            & $>0$ & $-b$ (const.)  & i-DMFT \\
$0<a\leq 1$    & $0$  & standard DFA   & standard hybrid functional \\
$0<a\leq 1$    & $>0$ & standard DFA   & thermally enhanced hybrid \\
\bottomrule
\end{tabular}
\end{table}

\section{Numerical method}
\label{sec:num-method}

In the previous section, we have derived the thermally enhanced hybrid functional for the fully interacting system at zero temperature. Following the generalized Kohn-Sham (GKS) formulation of hybrid schemes~\cite{Seidl1996,Goerling1997,Kuemmel2008}, we express it in terms of a set of orthonormal spin orbitals $\{\phi_i\}$ and their occupation numbers $\{f_i\}$, $0\leq f_i\leq 1$, which together define the 1RDM and the density of the reference system,
\begin{align}
    &\gamma(\rr\sigma,\rr'\sigma') = \sum_i f_i\,\phi_i(\rr\sigma)\,\phi_i^*(\rr'\sigma'),\\
    &\rho(\rr) = \sum_\sigma \gamma(\rr\sigma,\rr\sigma) = \sum_i f_i \sum_\sigma |\phi_i(\rr\sigma)|^2. \label{eq:dens}
\end{align}
Using the thermally enhanced hybrid functional from Eq.~\eqref{eq:Exctheta}, the approximated total interacting energy in terms of $(\{\phi_i\},\{f_i\})$ reads 
\begin{equation}\begin{aligned}\label{eq:energy-func}
    \tilde E &= F^{0,\theta}(\{\phi_i\},\{f_i\}) + \EH(\rho) + a\ExHFtheta(\{\phi_i\},\{f_i\}) \\
    &\quad+ (1-a)\ExDFA(\rho) + \EcDFA(\rho) + \int v(\rr)\rho(\rr)\rmd\rr.
\end{aligned}\end{equation}
We emphasize that the orbitals and occupation numbers above refer to those of the non-interacting reference system at temperature $\theta$ that reproduces exactly $\rho$.
The term $\ExHFtheta$ is just HF exchange, but with corresponding occupation numbers,
\begin{equation}
    \ExHFtheta = \frac{1}{2} \sum_{i,j} f_i f_j \bra{\phi_i\phi_j}\ket{\phi_i\phi_j}-\EH,
\end{equation}
where the two-electron integrals are understood to be antisymmetrized.
We also put in $F^{0,\theta}=T^{0,\theta}-\theta S^{0,\theta}$ from Eq.~\eqref{eq:functional-parts}, which evaluates on $(\{\phi_i\},\{f_i\})$ as
\begin{equation}
    F^{0,\theta}(\{\phi_i\},\{f_i\}) = \sum_i f_i \bra{\phi_i}\hat t\ket{\phi_i} -\theta\Sph(\{f_i\})
\end{equation}
with $\hat t$ the one-particle kinetic operator and the particle-hole symmetrized entropy
\begin{equation}
    \Sph(\{f_i\}) = -\sum_i \left( f_i\log(f_i) + (1-f_i)\log(1-f_i) \right). \label{eq:def-Sph}
\end{equation}
Here, we used the fact that the von Neumann entropy for a thermal but non-interacting ensemble evaluates as $\Sph$ known from the ideal Fermi gas. This fact even extends to all \emph{Gaussian fermionic states}~\cite{Mermin1963,BachLiebSolovej1994} (also called quasi-free states). The particle-hole symmetrized entropy is also closely connected to the \emph{nonfreeness} of states~\cite{Gottlieb2005,Gottlieb2007Properties} that was recently linked to correlation temperature~\cite{perspective}.

We are now in the position to optimize the total energy $\tilde E$ with respect to orbitals and their occupations.
In order to ensure the orthonormality of the spin orbitals and the conservation of the total particle number $N$, we introduce a Lagrangian with corresponding multipliers
\begin{equation}
    \mathcal{L}=\tilde E - \mu\bigg(\sum_i f_i - N\bigg) - \sum_{ij} \lambda_{ij}\left(\langle \phi_i|\phi_j\rangle - \delta_{ij}\right).
\end{equation}
We call the multiplier $\mu$ the chemical potential, while the diagonal values $\lambda_{ii}$ will be the orbital energies $\varepsilon_i = \lambda_{ii}/f_i$ in the canonical orbital representation, as in conventional KS-DFT. The minimization of the energy under the constraints, via the search for a stationary point of $\mathcal{L}$, requires two self-consistent solution steps. Variation with respect to the orbitals while keeping orbital occupations fixed yields the single-particle equations
\begin{equation}\label{eq:single-particle}
    \bigg( \hat h[\veff]+  
    \sum_jf_j(\hat J_j-a\hat K_j)  \bigg)\phi_i = \varepsilon_i \phi_i,
\end{equation}
where the one-particle Hamiltonian $\hat h[v]=\hat t+\hat v$ includes the local potential
\begin{equation}\label{eq:veff}
    \veff[\rho] = \vext +(1-a)\vxDFA[\rho] + \vcDFA[\rho].
\end{equation}
The overall method thus contains the Hartree and exchange operators of standard HF together with a DFA. The only difference to standard GKS is the occurrence of fractional occupations instead of a single Slater determinant. These occupations must in turn be determined by varying the Lagrangian with respect to the occupation numbers with fixed orbitals, leading to
\begin{equation}\label{eq:working-eq-frac-occ}
    \mu = \varepsilon_i - \theta\log\left(\frac{1-f_i}{f_i}\right) \Leftrightarrow f_i = \left(1+\rme^{(\varepsilon_i-\mu)/\theta}\right)^{-1}\!,
\end{equation}
which gives the Fermi-Dirac distribution. The chemical potential $\mu$ is here given by the condition $\sum_i f_i = N$.

These working equations are not entirely new; similar ones have been derived and used in previous works.~\cite{WangBaerends22-PRL,Irimia2023-correlation,Irimia2023-open-shell,Graf2026} However, these featured only the HF terms in Eq.~\eqref{eq:single-particle}, i.e., set $a=1$ and $\vxDFA = \vcDFA = 0$. These HF$\corrtemp$ approaches have been applied to compute dissociation curves of simple molecules such as hydrogen~\cite{WangBaerends22-PRL,Graf2026}, and the only additional parameter with respect to standard HF calculation was the correlation temperature $\theta$. Notable exceptions are \citet{Chai2017}, who already proposed a hybrid scheme on top of TAO-DFT, and \citet{Gibney2022b,Gibney2023,Gibney2024}, who combine DFT functionals embedded in RDMFT with a non-idempotency term that also allows a temperature interpretation.~\cite{perspective}

At this point, we must call attention to a distinctive feature of the thermally enhanced approach: the inclusion of an entropic term into our functionals can only correct energies downwards, compared to the zero-temperature SCF energy of the same functional. This can be seen simply from the fact that, for both HF and KS-DFT, the zero-temperature (aufbau) solution corresponds to an idempotent 1RDM, for which the entropy vanishes. It therefore remains in the search space of the thermally enhanced functional with unchanged energy, so the minimized energy can never exceed the zero-temperature SCF energy, for any value of $\theta$, including negative ones.

The only additional input compared to a standard DFT calculation is thus the temperature parameter $\theta$, and a substantial part of the present and previous research on the topic focuses on finding suitable values. Following a conjecture by \citet{Collins}, $\theta$ corresponds to the negative slope of the dependence of the cumulant energy $\Ecum$ (i.e., the exact energy minus the mean-field HF energy of the correlated ground state) with respect to the change in particle-hole symmetrized entropy ($\Sph$ from Eq.~\eqref{eq:def-Sph}) along a molecular distortion.~\cite{WangKnowles21-PRA,Graf2026,perspective} In this spirit, one might determine $\theta$ by performing a linear fit to the results of a high-level (e.g.~CASSCF) calculation via computing the one- and two-particle density matrices and from these the occupation numbers. However, absolute and dissociation energies obtained like this are typically not exact. Thus, it was often suggested to instead choose $\theta$ such that it exactly matches the reference dissociation energy. This method is also an important alternative in case of thermally enhanced DFT calculations, where it can be expected that the optimal values for $\theta$ differ significantly.~\cite{Lin2017} However, typical CASSCF calculations are based on HF orbitals and the HF Hamiltonian.~\cite{hupp03} Yet, some part of the cumulant energy that could be isolated from such calculations is already reproduced by the correlation functional but cannot be evaluated from the HF Hamiltonian, making it impossible to extract $\theta$ suitable for such calculations. As density functional approximations to use with the thermally enhanced method, we consider M06-HF~\cite{zhao06}, B3LYP~\cite{stephens94}, M06~\cite{zhao08}, and $\omega$B97X-V.~\cite{mardirossian14} With this selection, we want to test the performance of the method over a variety of hybrid density functionals, i.e., GGA functionals (B3LYP, $\omega$B97X-V), meta-GGA functionals (M06, M06-HF), including a range-separated functional with dispersion corrections ($\omega$B97X-V) and also one that uses 100\% exact exchange (M06-HF). The fractions of exact exchange in the other functionals are $a=0.2$ in B3LYP, and 0.27 in M06. In the range-separated $\omega$B97X-V functional, different proportions of exchange contributions are applied at different distances. This means that in the thermally enhanced hybrid functional from Eq.~\eqref{eq:Exctheta} we must replace the exchange part ($aE_x^\mathrm{HF\corrtemp}+(1-a)E_x^\mathrm{DFA}$) by
\begin{equation}
    a_\mathrm{sr}E_\mathrm{x,sr}^{\mathrm{HF}\corrtemp}+(1-a_\mathrm{sr})E_\mathrm{x,sr}^\mathrm{B97}+E_\mathrm{x,lr}^{\mathrm{HF}\corrtemp}.
\end{equation}
Here, ``lr'' and ``sr'' refer to functionals partitioned into long- and short-range contributions by means of the error function and we set $a_\mathrm{sr}=0.167$, corresponding to the usual value for $\omega$B97X-V~\cite{mardirossian14}.

A word on terminology for the discussed methods: We write ``[method]$\corrtemp$'' to an SCF method or hybrid functional if we replace the usual exact exchange part by $\ExHFtheta$ as in Eq.~\eqref{eq:Exctheta} on top of the non-interacting reference system $F^{0,\theta}$. As mentioned just before, ``HF$\corrtemp$'' is then exactly the ``i-DMFT'' method without the constant correlation correction $-b$. Further, ``DFT$\corrtemp$'' designates all DFT methods that include a thermally enhanced exact exchange part in form of a hybrid.

\section{Study of dissociation processes}
\label{sec:study-dissoc}

We now turn our attention to concrete molecular systems in dissociation processes.
As noted in Section~\ref{sec:num-method}, for each combination of molecule, SCF method, and employed basis set, we determine $\theta$ such that the dissociation energy matches reference values. In the case of hydrogen, we used the FCI calculation in the respective basis set as reference, while we rely on experimental dissociation energies for the larger nitrogen and ethylene. We also quantify the absolute energy shift relative to a reference calculation, $\Delta E_\mathrm{abs}^\mathrm{ref} = E^\mathrm{SCF\corrtemp} - E^\mathrm{ref}$, at the equilibrium distance. Notably, in the case of HF$\corrtemp$, this corresponds to an approximation to the parameter $b$ previously used as a global energy shift in the i-DMFT method.~\cite{WangBaerends22-PRL} In the cases of nitrogen and ethylene, we used single-reference B3LYP calculations in the respective basis set as reference values $E^\mathrm{ref}$, such that $\Delta E_\mathrm{abs}^\mathrm{B3LYP}$ quantifies the distance to the parent functional for B3LYP$\corrtemp$. The exact methodology for the determination of the dissociation energy, the equilibrium distance $r_\rme$, and the vibrational wave number $\tilde{\nu}$ is detailed in the Supplementary Material. The calculations were carried out using a modified version of the PySCF~\cite{Sun20} software package. The custom code to carry out thermally enhanced calculation can be found in the Supplementary Material.

\subsection{Results for hydrogen}
\label{sec:study-dissoc:hydrogen}

We first discuss the results for hydrogen, summarized in Table~\ref{tab:h2_idmft}.

\begin{table*}
\caption{Data from thermally enhanced calculations on hydrogen with different functionals with $\theta$ fitted to FCI dissociation energies of 0.165 $\mathrm{Ha}$ (cc-pVDZ) and 0.173 $\mathrm{Ha}$ (cc-pVTZ) in comparison to FCI results.}\label{tab:h2_idmft}
\begin{tabular}{llccrrrr}\toprule
Method & Basis & $n_\rme$ & $n_\mathrm{act}$ & $\theta/\mathrm{Ha}$ & $\Delta E_\mathrm{abs}^\mathrm{FCI}/\mathrm{Ha}$ & $r_\rme$ / Bohr & $\tilde{\nu}$ / cm$^{-1}$\\\midrule
FCI                         & cc-pVDZ & 2 & 10 & 0.09665$^\mathrm{a}$   & 0.04129$^\mathrm{a}$  & 1.440 & 4406 \\
HF$\corrtemp$               & cc-pVDZ & 2 & 10 & 0.09470                & 0.02870    & 1.437 & 4343\\
M06-HF$\corrtemp$           & cc-pVDZ & 2 & 10 & 0.08648                & $-0.00509$ & 1.445 & 4381\\
B3LYP$\corrtemp$            & cc-pVDZ & 2 & 10 & 0.04236                & $-0.01040$ & 1.445 & 4332\\
$\omega$B97X-V$\corrtemp$   & cc-pVDZ & 2 & 10 & 0.08164                & $-0.00817$ & 1.458 & 4239\\
M06$\corrtemp$              & cc-pVDZ & 2 & 10 & 0.04641                & $-0.00366$ & 1.437 & 4389\\\midrule

FCI                         & cc-pVTZ & 2 & 28 & 0.09429$^\mathrm{a}$   & 0.04703$^\mathrm{a}$  & 1.406 & 4411\\
HF$\corrtemp$               & cc-pVTZ & 2 & 28 & 0.09141                & 0.03305    & 1.409 & 4349\\
M06-HF$\corrtemp$           & cc-pVTZ & 2 & 28 & 0.08441                & $-0.00295$ & 1.411 & 4457\\
B3LYP$\corrtemp$            & cc-pVTZ & 2 & 28 & 0.04055                & $-0.00802$ & 1.409 & 4369\\
$\omega$B97X-V$\corrtemp$   & cc-pVTZ & 2 & 28 & 0.07918                & $-0.00488$ & 1.423 & 4274\\
M06$\corrtemp$              & cc-pVTZ & 2 & 28 & 0.04387                & $0.00072$  & 1.414 & 4259\\\midrule

Exp. & & & & & & 1.40~\cite{huber79} & 4401~\cite{huber79}\\\bottomrule
\multicolumn{8}{p{.65\linewidth}}{$^\mathrm{a}$ Values obtained from the least-squares fit of a linear curve to the $\Sph$--$\Ecum$ dependence along the dissociation.}\\
\end{tabular}
\end{table*}

The original i-DMFT method can reliably reproduce the equilibrium bond length and vibrational frequency computed at FCI level, especially in the larger basis set. However, significant shifts $\Delta E_\mathrm{abs}^\mathrm{FCI}$ emerge. We interpret the deviations in absolute energy values as a deficiency of the entropy functional in the regime of dynamic correlation, which is dominant near the equilibrium geometry. In i-DMFT, this shortcoming is mitigated in first order by introducing a constant offset.
Notably, the values for $\Delta E_\mathrm{abs}^\mathrm{FCI}$ when using DFT correlation functionals are smaller by a factor of 3 to 50 compared to HF; they are also negative in most cases, indicating a small double-counting of electron correlation. The M06 functional results in the smallest $|\Delta E_\mathrm{abs}^\mathrm{FCI}|$. The values for $\theta$ are also 8-55\% smaller when using DFT correlation functionals compared to the HF results. In particular, B3LYP results in the lowest $\theta$, followed by M06, then $\omega$B97X-V and then M06-HF. This ordering is consistent with the fraction of exact exchange in the functionals, as previously predicted by \citet{Lin2017} in the context of TAO-DFT. However, the values found here are very different from their estimate $\theta/\mathrm{Ha} = 0.007+0.052a$ in TAO-DFT, which would predict 0.0174 $\mathrm{Ha}$ for B3LYP, 0.021 $\mathrm{Ha}$ for M06, and 0.059~Ha for M06-HF.

With the DFT correlation functionals, the equilibrium distance and vibrational frequency have somewhat lower quality, especially with the $\omega$B97X-V functional, but most numerical results obtained with DFT$\corrtemp$ barely differ from HF$\corrtemp$. The functional with the closest match to FCI values in the cc-pVTZ basis is B3LYP. 

\begin{figure*}
    \centering
    \includegraphics[width=0.32\linewidth,clip,trim=0.45cm 0cm 1.6cm 1.4cm]{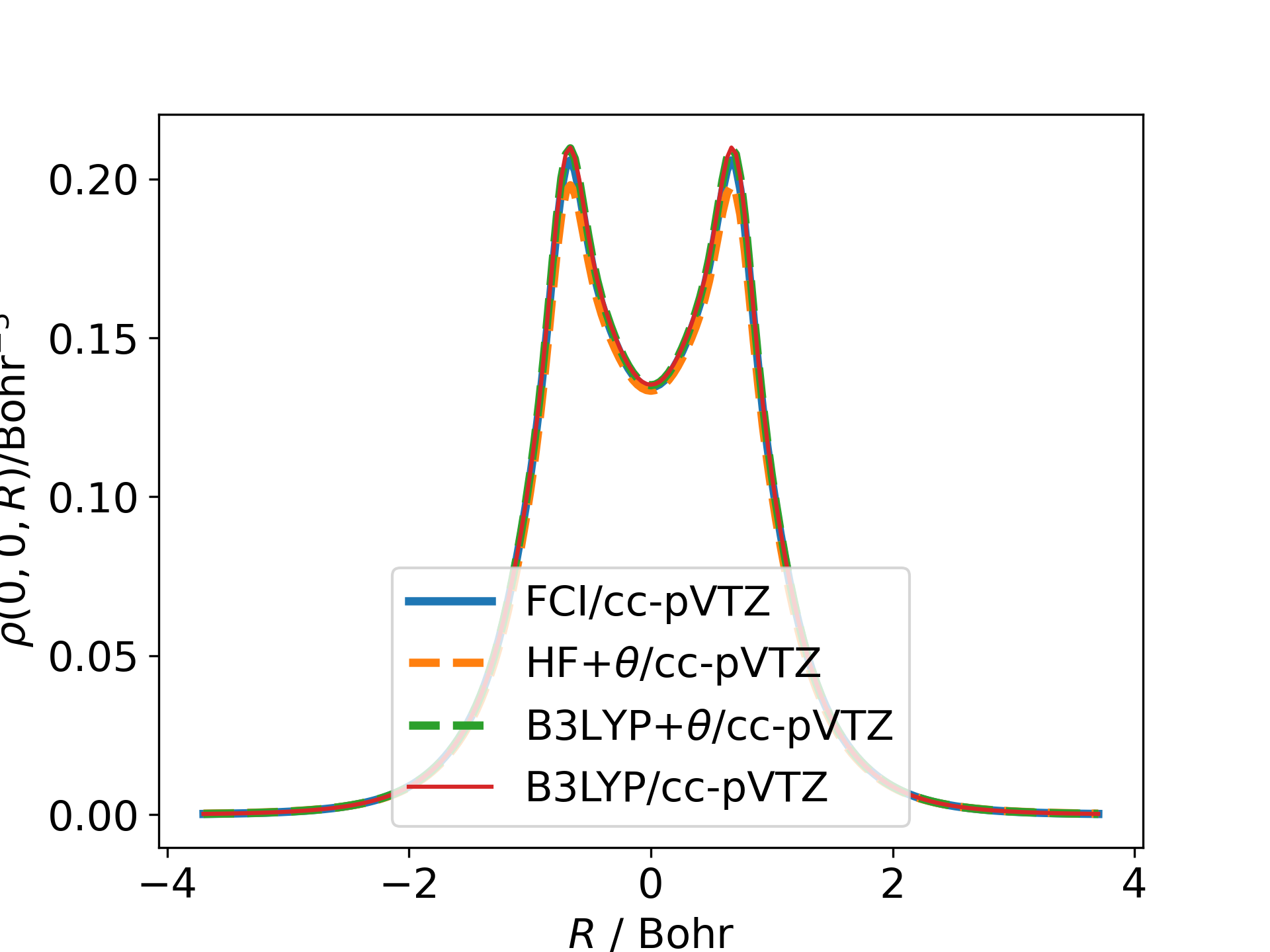}
    \includegraphics[width=0.32\linewidth,clip,trim=0.45cm 0cm 1.6cm 1.4cm]{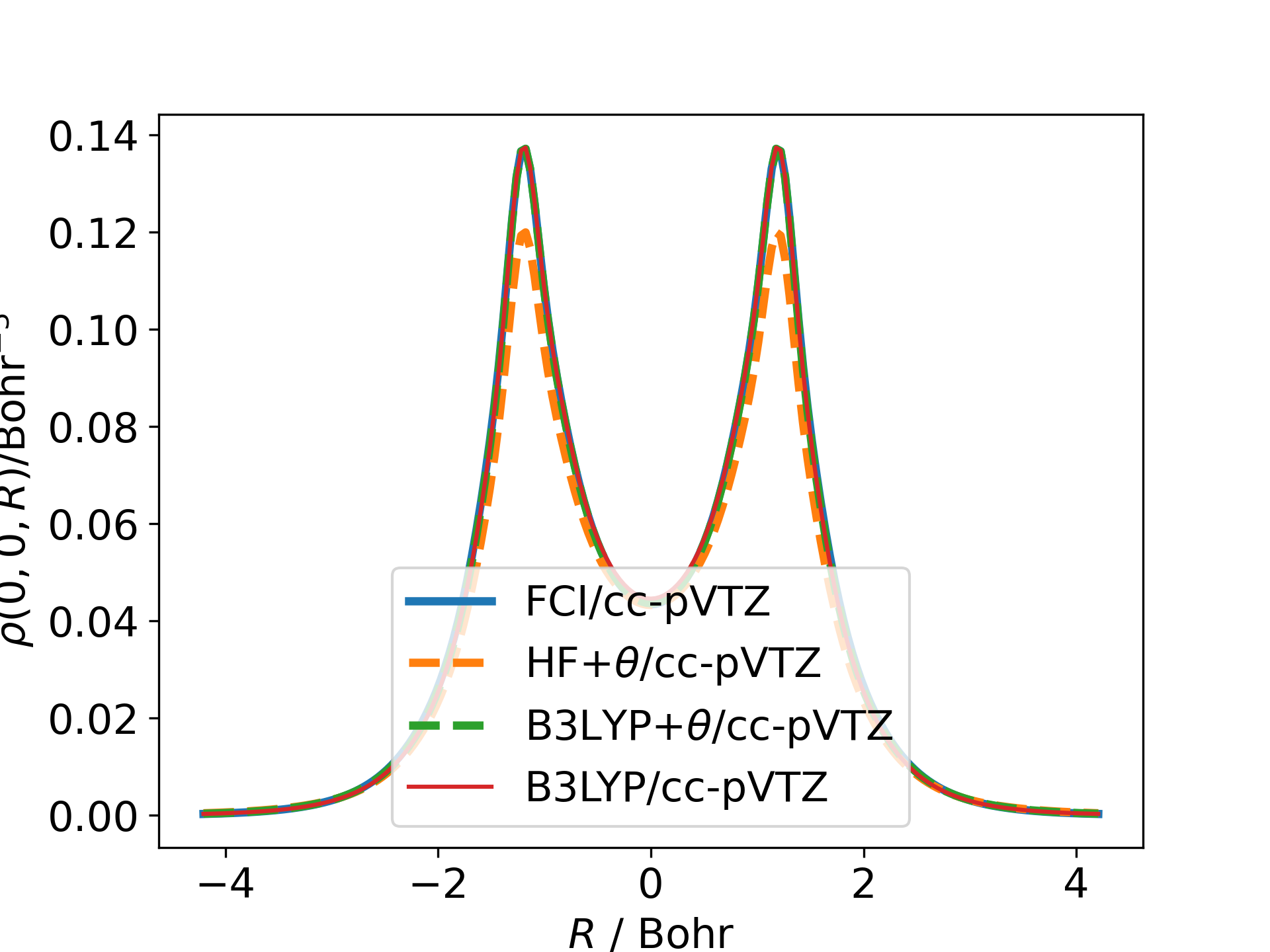}
    \includegraphics[width=0.32\linewidth,clip,trim=0.45cm 0cm 1.6cm 1.4cm]{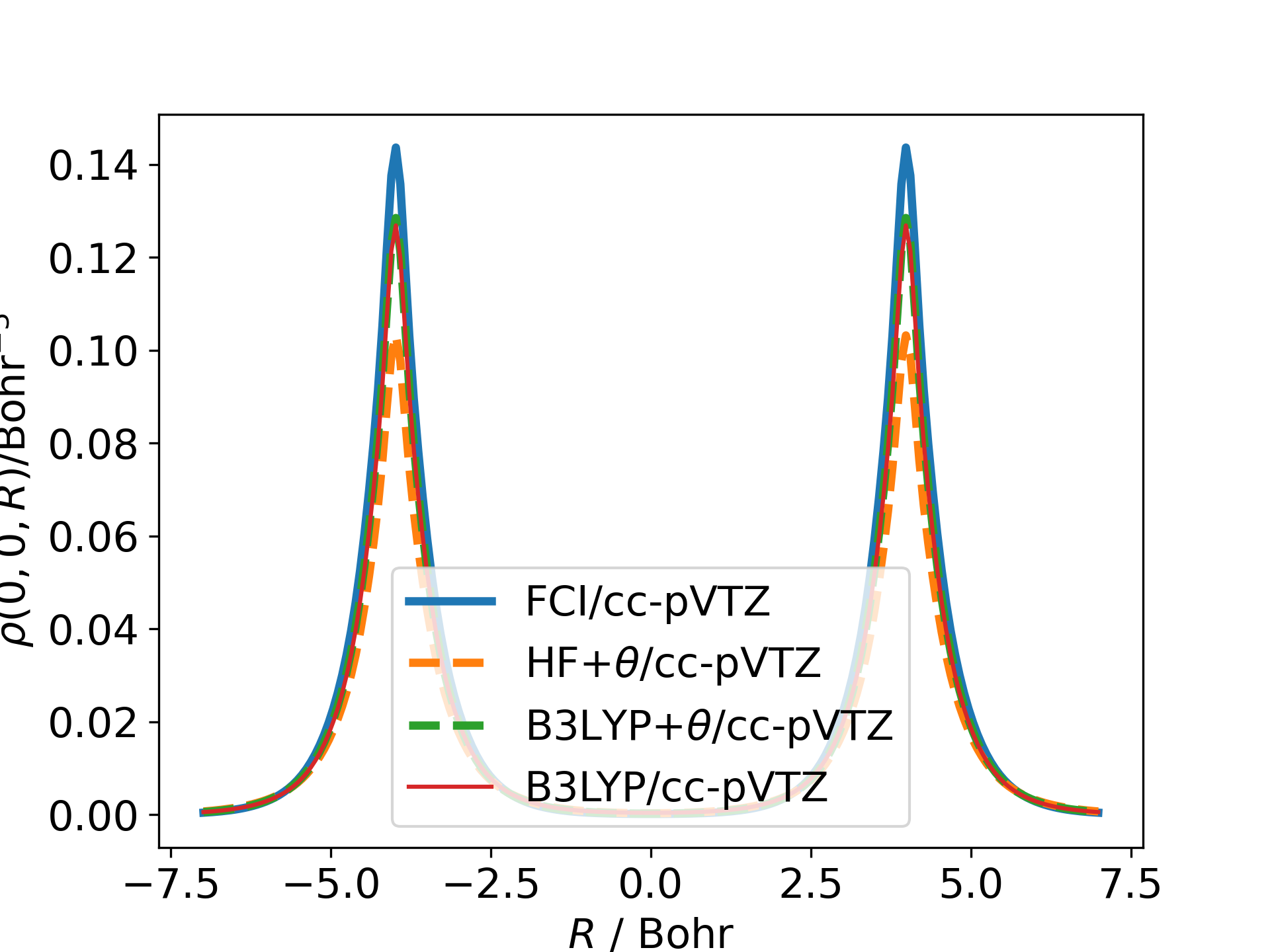}
    \caption{Electron density along the internuclear axis in the hydrogen molecule, at bond distances of 1.40 Bohr (left panel), 2.4 Bohr (middle panel), and 7.9 Bohr (right panel), computed with different methods.}\label{fig:h2-dens}
\end{figure*}

A straightforward conclusion from the change in $\Delta E_\mathrm{abs}^\mathrm{FCI}$ upon inclusion of correlation functionals is that this procedure allows one to explicitly include dynamic correlation within a thermally enhanced calculation, which otherwise appears as a constant shift. Since the energy shift does not affect the shape of the orbitals, but the form of the correlation functional does, a natural next question is whether it can correct the overly diffuse electron density obtained from HF$\corrtemp$.~\cite{Graf2026} In Fig.~\ref{fig:h2-dens}, we show the electron density along the internuclear axis for three different bond distances in the hydrogen molecule, computed with FCI, HF$\corrtemp$ and B3LYP$\corrtemp$.
While the HF$\corrtemp$ method results in too diffuse electron densities, especially at large bond distances, the correlation-corrected thermally enhanced calculations almost exactly reproduce the FCI results at bond distances of 1.4 and 2.4 Bohr, and reproduce the cusp at the nucleus significantly better at 7.9 Bohr. While small deviations remain, we can confirm that the additional treatment of dynamic correlation via a DFA improves the electron density.

\subsection{Results for nitrogen}
\label{sec:study-dissoc:nitrogen}

As an example for a system with multiple statically correlated electron pairs at dissociation, we consider the nitrogen molecule. Table~\ref{tab:n2_idmft} shows the fit results for different thermally enhanced calculations using the cc-pVDZ and cc-pVTZ basis sets.

\begin{table*}
\caption{Data from thermally enhanced calculations on nitrogen with different functionals with $\theta$ fitted to the experimental dissociation energy of 0.364 $\mathrm{Ha}$~\cite{lofthus77}, compared to CASSCF and B3LYP results.}\label{tab:n2_idmft}
\begin{tabular}{llccrrrr}\toprule
Method & Basis & $n_\rme$ & $n_\mathrm{act}$ & $\theta/\mathrm{Ha}$ & $\Delta E_\mathrm{abs}^\mathrm{B3LYP}/\mathrm{Ha}$ & $r_\rme$ / Bohr & $\tilde{\nu}$ / cm$^{-1}$\\\midrule
CASSCF                      & cc-pVDZ & 6  & 24 &        &           & 2.105 & 2368\\
B3LYP                       & cc-pVDZ & 14 & 28 & 0      &           & 2.087 & 2454\\
HF$\corrtemp$               & cc-pVDZ & 14 & 28 & 0.1173 & $0.5352$  & 2.076 & 2484\\
M06-HF$\corrtemp$           & cc-pVDZ & 14 & 28 & 0.1083 & $-0.0176$ & 2.091 & 2387\\
B3LYP$\corrtemp$            & cc-pVDZ & 14 & 28 & 0.0467 & $-0.0050$ & 2.096 & 2357\\
$\omega$B97X-V$\corrtemp$   & cc-pVDZ & 14 & 28 & 0.0893 & $0.0031$  & 2.117 & 2286\\
M06$\corrtemp$              & cc-pVDZ & 14 & 28 & 0.0522 & $0.0596$  & 2.100 & 2341\\\midrule

CASSCF                      & cc-pVTZ & 6  & 24 &        &           & 2.077 & 2387\\
B3LYP                       & cc-pVTZ & 14 & 60 & 0      &           & 2.062 & 2450\\
HF$\corrtemp$               & cc-pVTZ & 14 & 60 & 0.1175 & $0.5376$  & 2.057 & 2430\\
M06-HF$\corrtemp$           & cc-pVTZ & 14 & 60 & 0.1080 & $-0.0254$ & 2.070 & 2381\\
B3LYP$\corrtemp$            & cc-pVTZ & 14 & 60 & 0.0471 & $-0.0050$ & 2.071 & 2359\\
$\omega$B97X-V$\corrtemp$   & cc-pVTZ & 14 & 60 & 0.0895 & $0.0073$  & 2.095 & 2276\\
M06$\corrtemp$              & cc-pVTZ & 14 & 60 & 0.0528 & $0.0618$  & 2.072 & 2349\\\midrule

Exp. &  &  &  &  &  & 2.07~\cite{huber79} & 2359~\cite{irikura07}\\\bottomrule
\end{tabular}
\end{table*}

As for hydrogen, the HF$\corrtemp$ results lack a significant amount of correlation energy, which is reflected in the values of $\Delta E_\mathrm{abs}^\mathrm{B3LYP}$ being over 0.5~$\mathrm{Ha}$. Thermally enhanced DFT functionals, on the other hand, only show a small absolute energy deviation, and reproduce the absolute energy at equilibrium of the parent B3LYP functional without temperature well. The optimal values for $\theta$ are again lower with DFT functionals, but differ from those obtained for hydrogen, reproducing the previously observed variation of the electronic temperature from molecule to molecule~\cite{Graf2026,WangBaerends22-PRL,WangKnowles21-PRA} and showing a discrepancy to previous work on temperature-corrected DFT~\cite{Lin2017,Feng2025,Chen2022} where the electronic temperature was assumed identical for all systems. The smaller values do, however, have a similar proportion to each other. With a few percentage points of uncertainty, the DFT$\corrtemp$ optimal $\theta$ are 92\% (M06-HF), 42\% (B3LYP), 81\% ($\omega$B97X-V), and 47\% (M06) of the HF$\corrtemp$ values when averaging over the results for hydrogen and nitrogen.

DFT$\corrtemp$, with the exception of $\omega$B97X-V, generally reproduces bond lengths and the vibrational frequency with good quality, in particular more faithfully than KS-B3LYP and HF$\corrtemp$ calculations which significantly overestimate the vibrational frequencies. The thermally enhanced range-separated functional also underestimates bond strengths with both basis sets both compared to reference data and the experiment. For the other functionals, this is a sign that the DFT$\corrtemp$ approach is capable of describing the interplay of static and dynamic correlation along a potential energy surface for these systems. Again, B3LYP$\corrtemp$ emerges as the method with the best reproduction of experimental results. 

\begin{figure}
    \centering
    \includegraphics[width=\linewidth]{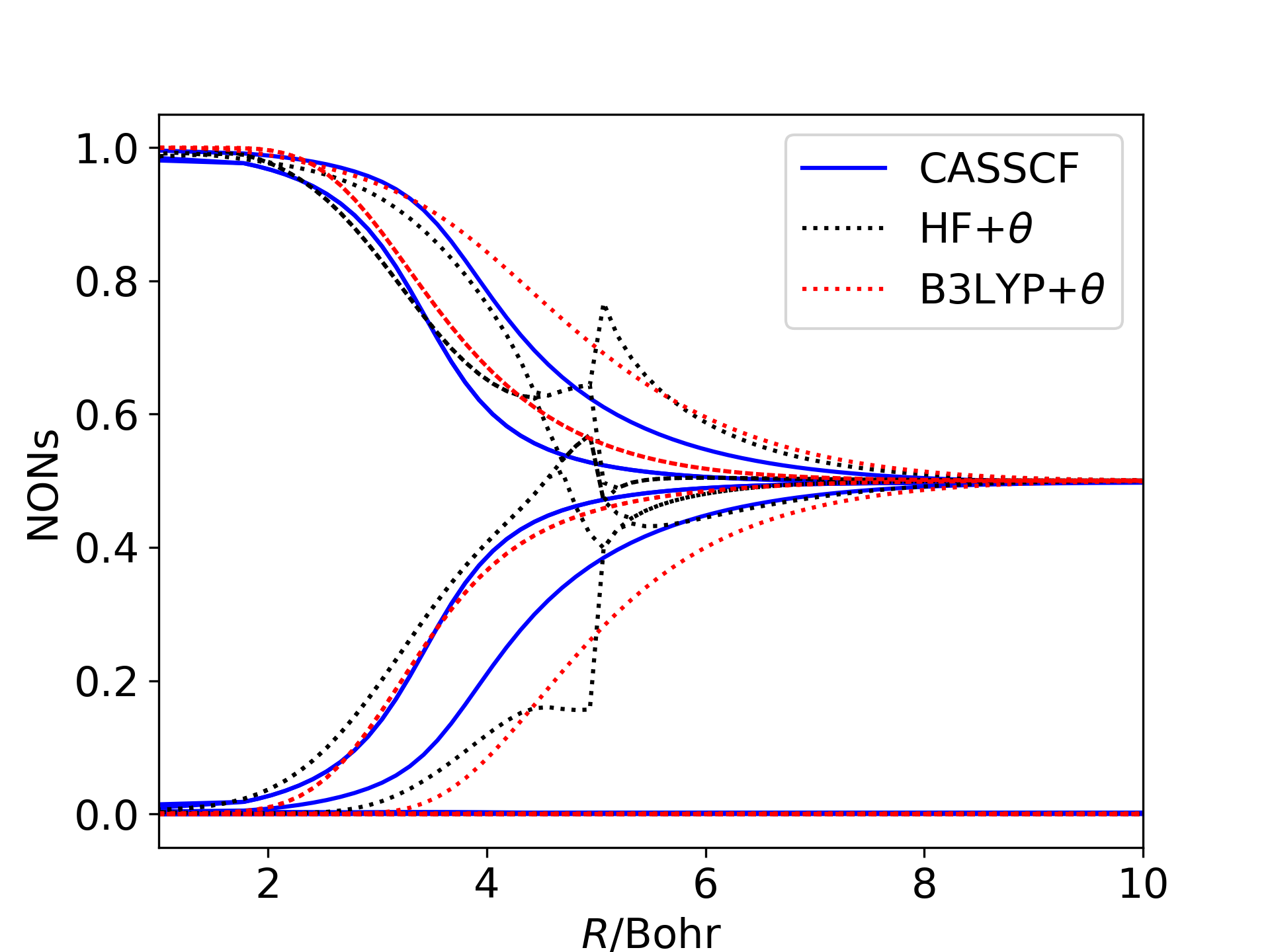}
    \caption{Natural occupation numbers of the orbitals 5-28 along the dissociation of nitrogen, computed with cc-pVDZ.}
    \label{fig:n2-nons}
\end{figure}

In our previous assessment of i-DMFT calculations on the nitrogen molecule~\cite{Graf2026}, we also compared the natural occupation numbers (NONs) with those obtained in CASSCF. We identified significant deviations, including an unphysical jump at a distance around 5 Bohr, as also reproduced in Fig.~\ref{fig:n2-nons}. The figure compares the NONs computed with CASSCF with those obtained with HF$\corrtemp$ and B3LYP$\corrtemp$ at the respective optimal $\theta$ with the cc-pVDZ basis set. The data shows that the inclusion of a correlation functional results in smoothly changing NONs over the entire dissociation, even though they still deviate from the CASSCF results. It should be noted that our theoretical approach does not enforce a match of the 1RDM or the NONs; thus, both natural orbitals and NONs may differ from reference wave function calculations. Yet, we still interpret the continuity and qualitative match of the NON pattern as a sign that the non-interacting thermal reference system is a better suited starting point for the adiabatic connection.

\subsection{Results for ethylene}
\label{sec:study-dissoc:ethylene}

\begin{figure}
    \centering
    \includegraphics[width=\linewidth]{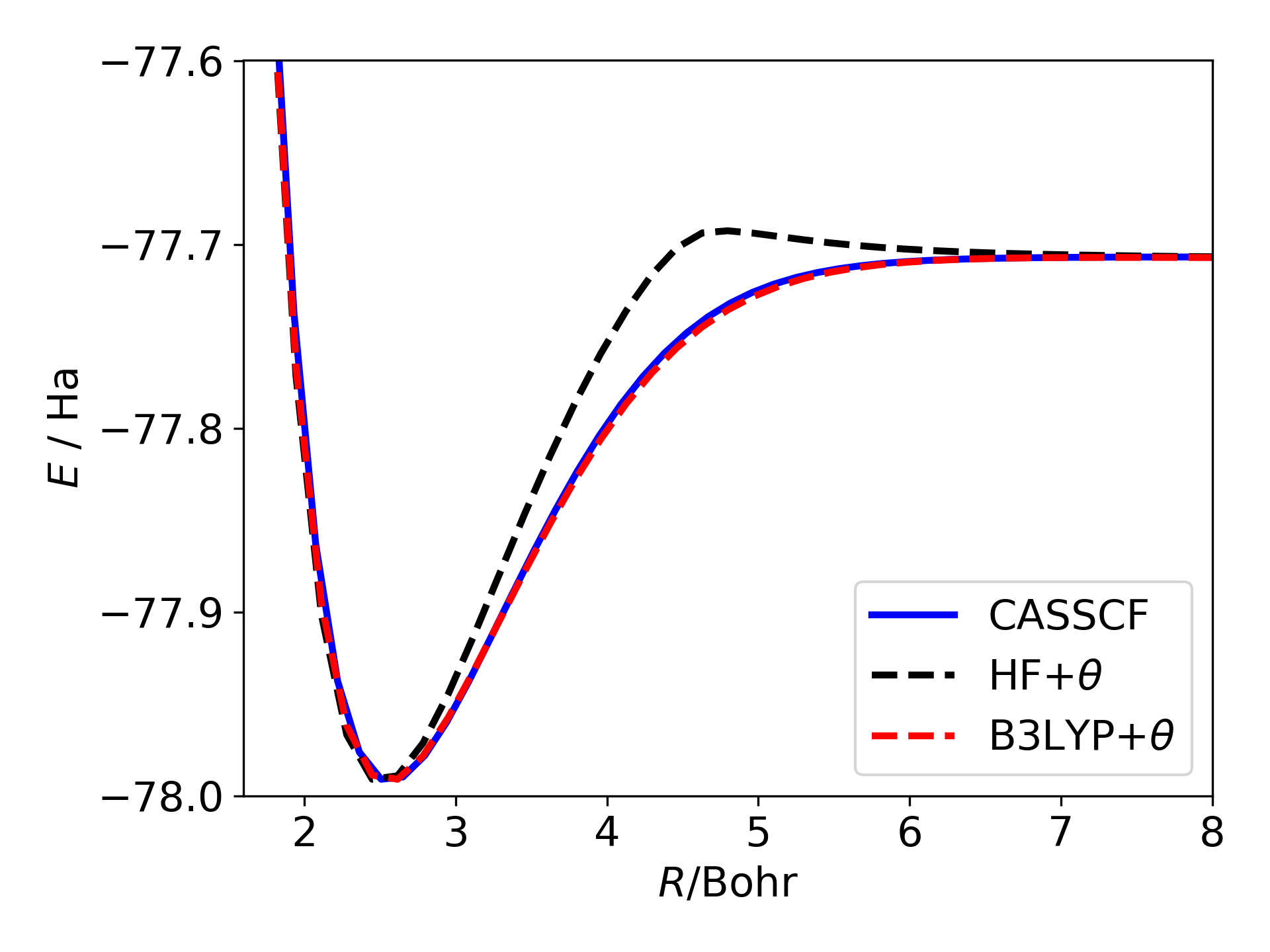}
    \caption{Symmetric dissociation curve for ethylene, computed with different methods and the STO-6G basis set, shifted to align the minimum energies.}\label{fig:c2h4-energies}
\end{figure}

\begin{figure}
    \centering
    \includegraphics[width=\linewidth]{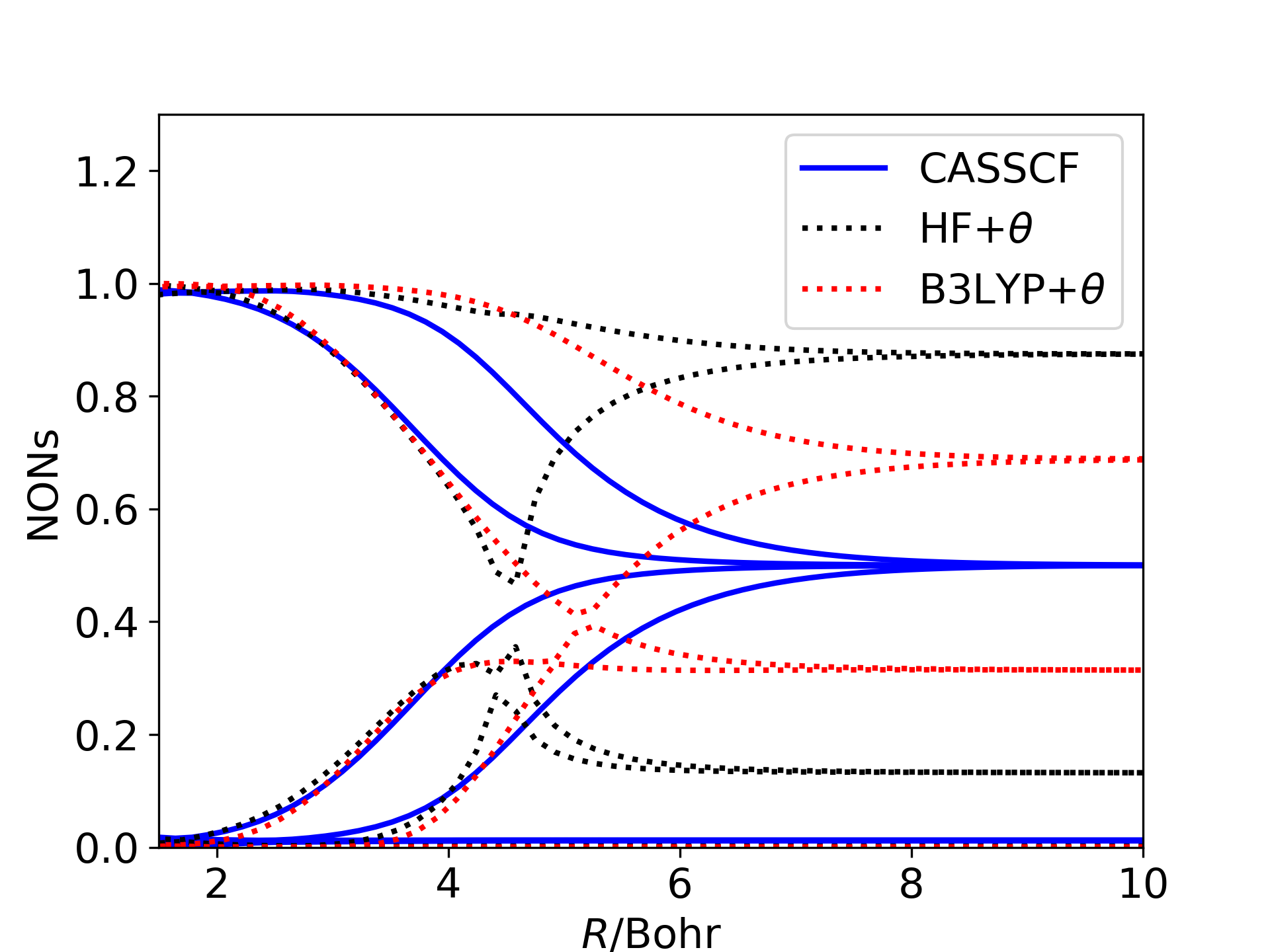}
    \caption{Natural occupation numbers along the dissociation of the C--C bond in ethylene, computed with different methods and the STO-6G basis set.}\label{fig:c2h4-nons}
\end{figure}

The final example we consider is the symmetric dissociation of the ethylene molecule. Previously, we showed that several issues arise with the HF$\corrtemp$ method.~\cite{Graf2026} Not only does an unphysical energy barrier emerge for this dissociation, but the NONs at dissociation actually resemble those of an excited state. To investigate if a density-based description of dynamic correlation can resolve these deviations, we carried out calculations with different functionals and basis sets. With each method, we computed 60 different D$_{2\mathrm{h}}$ geometries where the C--C bond length was varied from 0.1 to 10 Bohr, while the H--C bond lengths were kept at 2.054 Bohr and the H--C--H bond angles at 117.4$^\circ$. For each method, $\theta$ was determined by fitting the resulting dissociation energy to the experimental value -- it should be noted that a small inconsistency of a few mHa resulting from the computed dissociation into frozen fragments compared to a relaxed dissociation in the experiment can be expected. The numerical results are shown in Table~\ref{tab:c2h4_idmft}, and resulting potential energy surfaces and natural occupation numbers along the dissociation, compupted with CASSCF/STO-6G, HF$\corrtemp$/STO-6G, and B3LYP$\corrtemp$/STO-6G, in Figs.~\ref{fig:c2h4-energies} and~\ref{fig:c2h4-nons}.

\begin{table*}
\caption{Data from thermally enhanced calculations on ethylene with different functionals with $\theta$ fitted to the experimental dissociation energy of 0.286 $\mathrm{Ha}$~\cite{chase82,furtenbacher06,lerberghe72}, compared to CASSCF and B3LYP results.}\label{tab:c2h4_idmft}
\begin{tabular}{llccrrrr}\toprule
Method & Basis & $n_\rme$ & $n_\mathrm{act}$ & $\theta/\mathrm{Ha}$ & $\Delta E_\mathrm{abs}^\mathrm{B3LYP}/\mathrm{Ha}$ & $r_\rme$ / Bohr & $\tilde{\nu}$ / cm$^{-1}$\\\midrule
CASSCF                      & STO-6G  & 12 & 12  &        &           & 2.565 & 1571\\
B3LYP                       & STO-6G  & 16 & 14  & 0      &           & 2.457 & 1658\\
HF$\corrtemp$               & STO-6G  & 16 & 14  & 0.1049 & $0.5317$  & 2.515 & 1698\\
M06-HF$\corrtemp$           & STO-6G  & 16 & 14  & 0.1072 & $0.0084$  & 2.550 & 1570\\
B3LYP$\corrtemp$            & STO-6G  & 16 & 14  & 0.0490 & $-0.0079$ & 2.552 & 1550\\
$\omega$B97X-V$\corrtemp$   & STO-6G  & 16 & 14  & 0.0911 & $-0.0051$ & 2.559 & 1566\\
M06$\corrtemp$              & STO-6G  & 16 & 14  & 0.0530 & $0.0610$  & 2.547 & 1560\\\midrule

B3LYP                       & cc-pVDZ & 16 & 48  & 0      &           & 2.458 & 1550\\
HF$\corrtemp$               & cc-pVDZ & 16 & 48  & 0.0507 & $0.5490$  & 2.504 & 1670\\
M06-HF$\corrtemp$           & cc-pVDZ & 16 & 48  & 0.0767 & $0.0108$  & 2.542 & 1508\\
B3LYP$\corrtemp$            & cc-pVDZ & 16 & 48  & 0.0316 & $-0.0017$ & 2.531 & 1523\\
$\omega$B97X-V$\corrtemp$   & cc-pVDZ & 16 & 48  & 0.0666 & $0.0169$  & 2.541 & 1527\\
M06$\corrtemp$              & cc-pVDZ & 16 & 48  & 0.0338 & $0.0750$  & 2.522 & 1553\\\midrule

B3LYP                       & cc-pVTZ & 16 & 116 & 0      &           & 2.440 & 1535\\
HF$\corrtemp$               & cc-pVTZ & 16 & 116 & 0.0486 & $0.5570$  & 2.490 & 1649\\
M06-HF$\corrtemp$           & cc-pVTZ & 16 & 116 & 0.0749 & $0.0044$  & 2.525 & 1511\\
B3LYP$\corrtemp$            & cc-pVTZ & 16 & 116 & 0.0309 & $-0.0021$ & 2.512 & 1525\\
$\omega$B97X-V$\corrtemp$   & cc-pVTZ & 16 & 116 & 0.0657 & $0.0212$  & 2.524 & 1526\\
M06$\corrtemp$              & cc-pVTZ & 16 & 116 & 0.0334 & $0.0751$  & 2.502 & 1543\\\midrule

Exp. &  &  &  &  &  & 2.530~\cite{herzberg66} & 1623~\cite{lerberghe72}\\\bottomrule
\end{tabular}
\end{table*}

An important finding is that B3LYP$\corrtemp$ computes a dissociation curve extremely close to the CASSCF results: the unphysical dissociation barrier obtained with HF$\corrtemp$ vanishes. This is a first clear sign for the qualitative superiority of the DFA-based methods and can be further corroborated by evaluating the NONs along the dissociation. Both HF$\corrtemp$ and B3LYP$\corrtemp$ follow the CASSCF reference data well until ca. 4 Bohr, but approach different dissociation limits afterwards. Instead of the occupation numbers of the 8 strongly correlated spin orbitals all becoming 0.5, they approach values around 0.87 and 0.13 with the HF$\corrtemp$ method and 0.69 and 0.31 with B3LYP$\corrtemp$. The occupations found with HF$\corrtemp$ have previously been associated with an excited state of the dissociated molecule~\cite{Cheung1979,Graf2026}, which can explain the dissociation barrier seen in HF$\corrtemp$. With B3LYP$\corrtemp$ method, we find that the occupations in the dissociation limit are more strongly mixed, tracking the characteristic occupation pattern of the fully correlated reference more closely.

Returning to the numerical results in Tab.~\ref{tab:c2h4_idmft}, a difference to the previous examples is that two of the functionals, $\omega$B97X-V and M06-HF, lead to larger optimal values for $\theta$ than the original HF functional in the cc-pVDZ and cc-pVTZ bases. Other DFAs still lead to a reduction in $\theta$ and all show much smaller $\Delta E_\mathrm{abs}^\mathrm{B3LYP}$ than the HF$\corrtemp$ method. A possible reason for $\theta$ with HF$\corrtemp$ being smaller is that it results from the dissociation into an excited state, while true ground state would require a larger $\theta$, while the DFT-based methods dissociate into the correct electronic state, thus resulting in a $\theta$ closer to the one observed for the previous examples. 

In general, bond lengths are predicted slightly larger and vibrational frequencies slightly lower when using a DFT$\corrtemp$ variant compared to the HF$\corrtemp$ or KS-B3LYP results. For the bond lengths, this results in numbers much closer to the experiment and the CASSCF reference in the STO-6G basis set, while the situation is less clear for the vibrational frequencies. While in the STO-6G basis set, the thermally enhanced DFT methods again reproduce the CASSCF value much better, none was computed for larger basis sets. It may appear that HF$\corrtemp$ is matching the experimental value best, but a mismatch might also result from the fact that we did not evaluate the vibrational frequency along the true eigenmode but kept the CH$_2$ fragments frozen along the dissociation. 

Based on the better match with CASSCF reference data where available, the better match with the equilibrium bond distance and the qualitatively correct reproduction of the barrierless dissociation surface, we still conclude that the DFT$\corrtemp$ approaches are superior to established KS-DFT and i-DMFT methods for the description of small dissociating molecules.

\section{Examination of the role of the correlation temperature}
\label{sec:theta-approaches}

From the numerical results presented in Section~\ref{sec:study-dissoc} and elsewhere~\cite{WangBaerends22-PRL,Graf2026}, it is evident that the value of the correlation temperature has a direct effect on the quality of thermally enhanced SCF calculations. Determining a suitable $\theta$, therefore, represents the fundamental task when employing any method from the SCF$\corrtemp$ family. This is particularly challenging since the parameter lacks a clear physical interpretation in the adiabatic connection: while $\theta$ is handled as a physical temperature of our chosen reference, we must emphasize that the reference itself is purely \emph{fictitious}. In principle, $\theta$ serves a solely auxiliary purpose and does not appear in the exact density functional. If an exact treatment of the framework established in Section~\ref{sec:thermal-hybrid} were feasible, we would find any temperature of the non-interacting reference fully compensated by the integral performed during the adiabatic connection (Eq.~\eqref{eq:AC}).
Practically speaking, however, the exact evaluation of the integral is intractable, such that our approach recovers the thermally enhanced hybrid approximations from Eq.~\eqref{eq:dens-func-interacting}, where the entropy contribution $-\theta S^{0,\theta}$ from the reference in $F^{0,\theta}$ is still present. Since the lack of a prescribed meaning in the context of the interacting target system leaves $\theta$ as a tunable parameter, it appears reasonable to interpret its role solely pragmatically as a means to improve the approximation, e. g., by tuning it to match the energy of a reference calculation. However, the quest for a reliable ab initio method to compute $\theta$ requires a more sophisticated understanding of the parameter's meaning in the context of the fully interacting physical system (at zero temperature).

In this section, we test various approaches for obtaining values of $\theta$ and identify which methods recover a correlation temperature that yields reasonable agreement between the energy surfaces of SCF$\corrtemp$ methods and a high-level reference. For this, we benchmark the resulting correlation-temperature values against geometry-specific sets of $\theta$ that reproduce the exact absolute energy surface (see below). We quantify the success of each approach based on how well it aligns with this benchmark.
The methodologies for determining $\theta$ that have emerged so far are relatively diverse in cost and complexity, but, most importantly, in their proposed universality.
Several approaches have focused on optimizing $\theta$ by minimizing the error of a relative energy in SCF$\corrtemp$ calculations, as also done in Section~\ref{sec:study-dissoc}.
In the context of TAO-DFT~\cite{Chai2012}, results for individual molecules during deformation processes as well as for entire benchmark sets led to an optimal global $\theta$ of 7~mHa for single-reference systems when combined with a pure LDA functional; however, for multi-reference systems the values vary strongly (between 5 and 50~mHa). We emphasize that the TAO-DFT method differs from ours due to the inclusion of the $E_\theta$ term, which was discussed in Section~\ref{sec:dft-thdyn}. The term vanishes in our method due to the approximation at the endpoints of the chosen adiabatic connection path. Also, we do not expect a system-independent value for $\theta$ in our framework, based on the results in Section~\ref{sec:study-dissoc} and a previous assessment~\cite{Graf2026}.

Besides approaches focused on deformations or reactivity of one molecule or a set, others are able to determine $\theta$ for specific geometries. In one approach, the parameter is chosen such that a system's ionization and attachment energies match its orbital energies~\cite{Lin2017}; another one fitted it such that the TAO-DFT calculation reproduces the occupation number of the LUMO from a reference calculation.~\cite{Chen2022} While the introduction of a geometry-specific $\theta$ is a different approach from that taken in previous works on i-DMFT, it is in line with a previous argument that requires $\theta$ to vary, e.g., as a nonconstant function of the entropy.~\cite{perspective}

In this work, we compare some of these and newly devised approaches for the determination of $\theta$. For this, we first recognize that the role of $\theta$ in SCF$\corrtemp$ is twofold: on the one hand, it directly affects the SCF$\corrtemp$ energy by scaling the contribution of the entropic term, while on the other hand, it controls the Fermi-Dirac distribution of the fractional occupations. Consequently, the methods considered here are split into three categories:

\begin{enumerate}[(A)]
    \item\label{cat1} Energy-based methods that seek a correlation temperature to match a reference in terms of either its dissociation energy with a system-specific $\theta$ (A1), or the entire absolute energy surface with geometry-specific values (A2).
    \item\label{cat2} Occupation-based methods that seek geometry-specific correlation temperatures by minimizing the deviation of an SCF$\corrtemp$ calculation to reference NONs (B1), or a reference entropy (B2).
    \item\label{cat3} Occupation-based methods which take $\theta$ as the temperature of a Fermi-Dirac distribution fitted to the NONs of a reference calculation as a function of specific orbital energies. We explore orbital energies evaluated from a high-level calculation (C1), from mean-field SCF calculations at zero temperature (C2), or from SCF$\corrtemp$ calculations (C3).
\end{enumerate}

These approaches are detailed in Sections~\ref{sec:theta-en-op}, \ref{sec:theta-match-op} and \ref{sec:theta-occ-op}, respectively. They are applied to an instructive example of a system featuring strong static correlation: the dissociation of nitrogen (14 electrons), which we evaluate in the cc-pVDZ basis (28 orbitals). For methods that require high-level reference calculations, we performed CASSCF(6,24)/cc-pVDZ calculations. 
Our results are presented in Fig.~\ref{fig:n2_kappa} and discussed in Section~\ref{sec:theta-op-results}. We performed the same series of calculations both for HF$\corrtemp$ (upper row, Fig.~\ref{fig:n2_kappa}) and B3LYP$\corrtemp$ (bottom row). Additional results on torsion in the ethylene molecule are provided in the Supplementary Material.

\subsection{Energy-based optimization of SCF+\texorpdfstring{$\theta$}{theta} calculations}\label{sec:theta-en-op}

For a global optimization of $\theta$ in method family (A1), we choose its value in an SCF$\corrtemp$ method such that a chosen reference dissociation energy $D_\rme$ is recovered, as previously suggested for i-DMFT~\cite{WangBaerends22-PRL} and already applied in the previous section. Details on the computation of the dissociation energy can be found in the Supplementary Material.
In other words, this ansatz represents an interpolation between two characteristic points of the dissociation process: the equilibrium bond length and the dissociation limit. The dissociation energies to be matched may be taken from any source; here, we compare the results for dissociation energies from three different methods.

Besides computing this energy difference with CASSCF, we propose to approximate the dissociation energy at SCF level with zero-temperature HF or DFT calculations by subtracting the ground-state energy of the system at the equilibrium geometry ($^1\Sigma_g$ state of nitrogen) from the energies of the separate fragments (two nitrogen atoms in the $^4$S$_{3/2}$ state, evaluated with restricted-open-shell HF/KS calculations). This yields reasonable approximations to the dissociation energy with a single-reference method, despite the underlying methods not being well-suited to the setting of strong static correlation, and, unlike the other approaches discussed here, allows one to compute $\theta$ and thus the dissociation curve entirely at mean-field cost.

The results for the dissociation energies computed with the different reference methods and the computed values of $\theta$ are listed in Table~\ref{tab:e_diss_n2_theta_glob}. 

\begin{table}
\centering
\caption{Dissociation energies and globally interpolated values of $\theta$ in nitrogen in the cc-pVDZ basis set. Dissociation energies were computed via CASSCF, restricted HF, and restricted KS-B3LYP (see text for details). The results are compared with the experimental results.}
\label{tab:e_diss_n2_theta_glob}
\begin{tabular*}{\columnwidth}{@{\extracolsep{\fill}}lc|cc@{}}
\toprule
reference method & $D_\rme$ & $\theta$ & $\theta$\\
for diss. energy & & (HF$\corrtemp$) & (B3LYP$\corrtemp$)\\
\midrule
CASSCF(6,24)                   & 0.330~Ha & 0.1225~Ha & 0.0513~Ha \\
\hspace{10pt} relative to exp. & 90.6\%   & 104.4\%   & 109.8\%   \\ \addlinespace
HF                             & 0.178~Ha & 0.1482~Ha &     --    \\
\hspace{10pt} relative to exp. & 48.9\%   & 126.3\%   &           \\ \addlinespace
B3LYP                          & 0.357~Ha &     --    & 0.0476~Ha \\
\hspace{10pt} relative to exp. & 98.0\%   &           & 101.9\%   \\
\midrule
exp.                           & 0.364~Ha~\cite{lofthus77} & 0.1173~Ha & 0.0467~Ha\\\bottomrule
\end{tabular*}
\end{table}

As expected, the approximations to the dissociation energy vary widely across methods. Both CASSCF(6,24) and B3LYP reproduce the experimental value relatively well, but the restricted HF ansatz performs worse, deviating from the experimental $D_\rme$ by a factor of more than 2. Notably, B3LYP approximates the experimental value even better than CASSCF.
The inverse trend results for the values of the globally fitted $\theta$, since the optimized value increases with a decrease in $D_\rme$ for a specific mean-field functional. Thus, especially for HF$\corrtemp$, values of $\theta$ vary significantly, while there is good agreement between experiment, CAS reference, and B3LYP reference for B3LYP$\corrtemp$. While DFT calculations cannot be assumed to be competitive against high-level methods in every setting, this example shows that computing $D_\rme$ via KS-DFT calculations on fragments can be successful.

\begin{figure*}[!t]
    \includegraphics[width=\linewidth]{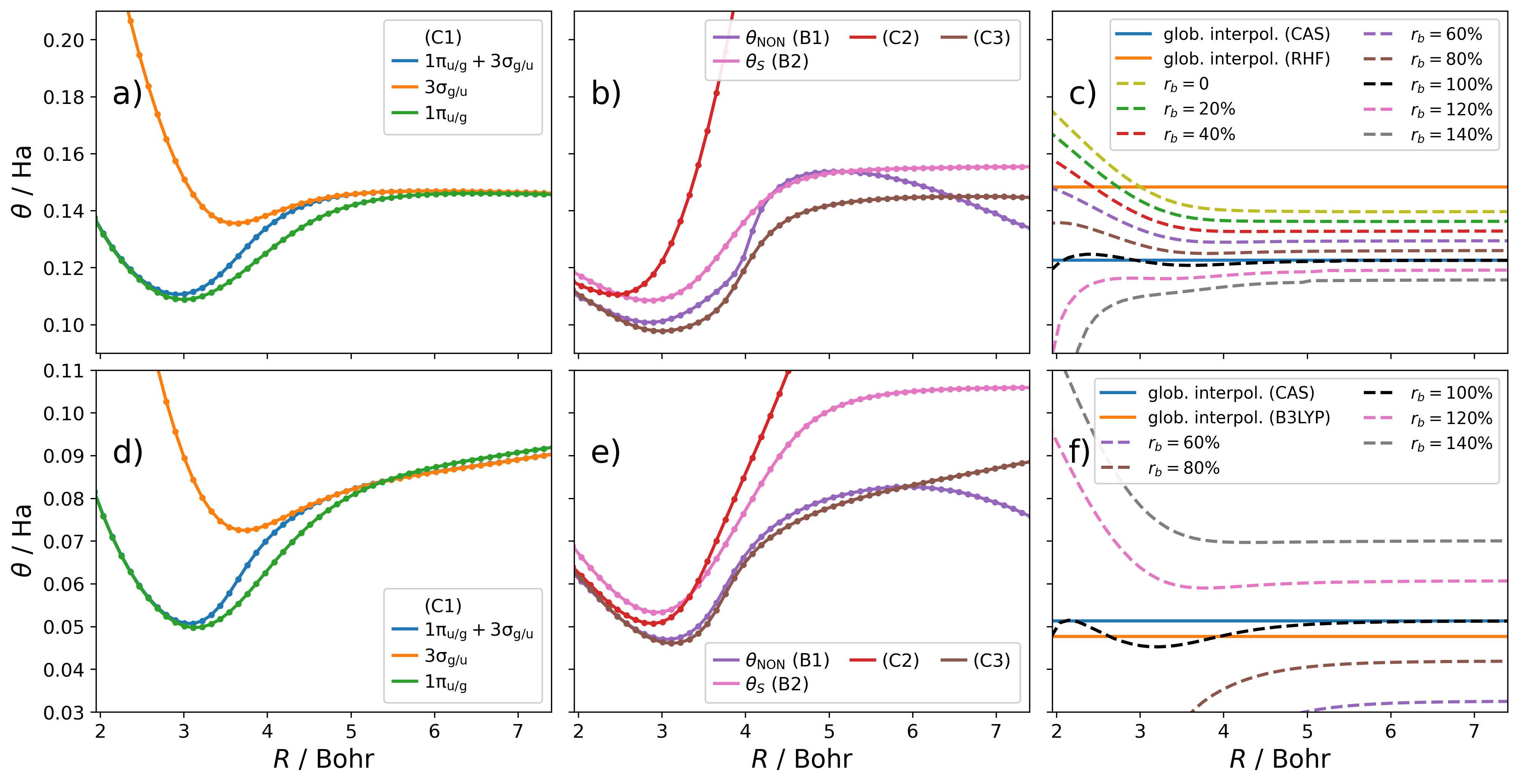}
    \caption{Correlation temperatures $\theta$ along the dissociation of nitrogen evaluated in the cc-pVDZ basis, with CASSCF(6,24)/cc-pVDZ as reference. Top row (a--c): HF$\corrtemp$; bottom row (d--f): B3LYP$\corrtemp$.
    Left column (a, d): (C1) $\theta$ from fits to the reference NONs using orbital energies from the reference, for the six statically correlated orbitals ($1\uppi_{u/g} + 3\upsigma_{g/u}$) or for the $3\upsigma_{g/u}$ or $1\uppi_{u/g}$ orbitals only.
    Middle column (b, e): $\theta$ chosen such that SCF$\corrtemp$ best approximates all reference NONs (``$\theta_\mathrm{NON}$ (B1)'') or the reference entropy (``$\theta_S$ (B2)''); $\theta$ from fits to the reference NONs of six orbitals, using orbital energies from a zero-temperature SCF calculation (C2) or from a SCF$\corrtemp$ calculation using $\theta_0=\theta_S$ (C3).
    Right column (c, f): $\theta$ from fits to energies. Solid lines: (A1) with global $\theta$ reproducing the dissociation energy from the reference or from zero-temperature RHF (c) or B3LYP (f) calculations. Dashed lines: (A2) with geometry-specific $\theta$ reproducing the absolute reference energy at each $R$ for a functional shifted by the constant $-b$, with $r_b = b/\Delta E^\mathrm{abs,CAS}_{\mathrm{SCF}\corrtemp}$ with $\Delta E^\mathrm{abs,CAS}_{\mathrm{SCF}\corrtemp}$ as defined in Sec.~\ref{sec:study-dissoc}.}
    \label{fig:n2_kappa}
\end{figure*}

The globally optimized values of $\theta$ are visualized as solid horizontal lines in panels c and f of Fig.~\ref{fig:n2_kappa} (``glob. interpol.'').

In method family (A2), we carry out geometry-specific optimizations of $\theta$ that recover the CAS reference energy at each point along the dissociation curve. The resulting values are plotted as dashed lines in panels c and f of Fig.~\ref{fig:n2_kappa}.
Crucially, the optimization of $\theta$ in this method now depends on the absolute energy obtained via our thermally enhanced functional. In Section~\ref{sec:study-dissoc}, we showed that benchmarking SCF$\corrtemp$ results against a CAS reference can leave a significant offset, $\Delta E_{\mathrm{SCF}\corrtemp}^\mathrm{abs,CAS}$, when $\theta$ is chosen to match the dissociation energy. 
Consequently, fitting the energy separately at each point along the dissociation results in significant fluctuations in $\theta$, especially near the equilibrium geometry where $S$ is low. To overcome this problem, we reintroduce the constant offset $-b$ (beware the negative sign) into the functional, as it appears in i-DMFT. The panels in the right column of Fig.~\ref{fig:n2_kappa} cover results for different values of $b$. We express it relative to $\Delta E_{\mathrm{SCF}\corrtemp}^\mathrm{abs,CAS}$, where $\Delta E^\mathrm{abs,CAS}_{\mathrm{HF}\corrtemp}=0.143$~Ha and $\Delta E^\mathrm{abs,CAS}_{\mathrm{B3LYP}\corrtemp}=-0.390$~Ha. 
Note that the value obtained for B3LYP$\corrtemp$ is negative, i.e., the energy curve obtained with the globally fitted value, $\theta=0.0513$~Ha, lies below the wave-function reference at the equilibrium geometry. On this note, we recall that the inclusion of an entropic term in our thermally enhanced approach can only correct energies downwards from the zero-temperature energy of the same functional. This was discussed in more detail in Section~\ref{sec:num-method}.
In other words, $\theta$ can only be optimized at those points where the zero-temperature SCF energy of the respective functional lies above the reference. Therefore, in the case of B3LYP$\corrtemp$, the offset $b$ serves the additional purpose of shifting the functional such that a $\theta$ recovering the reference energy can be obtained in the first place.

\subsection{Occupation-based optimization of SCF+\texorpdfstring{$\theta$}{theta} calculations}\label{sec:theta-match-op}

With the second class of methods, we turn away from the energy as the target quantity. Instead, we optimize $\theta$ in an SCF$\corrtemp$ calculation to approximate the fractional occupation number spectrum of a high-level reference. As the reference, we use again the results from CASSCF(6,24)/cc-pVDZ. Importantly, based on their role in density functional theory, we understand that the minimizer (orbitals and occupations) of the fictitious non-interacting system need not recover the true ground state beyond its density. Nevertheless, previous i-DMFT studies~\cite{Graf2026} and our DFT$\corrtemp$ results in Section~\ref{sec:study-dissoc} indicate that certain characteristics of the true ground-state spectrum can still be recovered, which motivates the methods discussed in the following. 
Since our results suggest a good approximation of all occupations (strongly occupied core, fractionally occupied valence, and weakly occupied virtual), we let the sum in Eq.~\ref{eq:ell2-error-nons} run over all orbitals. Fits of a subset including only the strongly correlated HOMO-2 to LUMO+2 orbitals (as discussed in the next section) yield nearly identical results.

In (B1), we evaluate $\theta$ by minimizing the deviation from reference NONs, which we quantify via the $\ell^2$-error:
\begin{equation}\label{eq:ell2-error-nons}
    \theta_\mathrm{NON} = \argmin_\theta\overbrace{\sum_i\Bigg[\underbrace{\left(1+\rme^{(\varepsilon_i(\theta)-\mu(\theta)/\theta}\right)^{-1}}_{=f_i\text{ (see Eq.~\eqref{eq:working-eq-frac-occ})}}-n_i^\mathrm{ref}\Bigg]^2}^{\ell^2\text{-error}}.
\end{equation}
Here, both reference NONs and fractional SCF$\corrtemp$ occupations are ordered and matched by size.
In this, we recognize that neither $\mu$ nor the orbital energies are free in the SCF$\corrtemp$ methods: $\{\varepsilon_i\}$ depend on $\theta$ via the working equations discussed in Section~\ref{sec:num-method}, while $\mu$ is bound to reproduce the total number of electrons. 

The maximum deviation found along the dissociation curve (in terms of the root-mean-square error over all 56 spin orbitals in the cc-pVDZ basis set for N$_2$) in the occupations of $0.020$ (HF$\corrtemp$) and $0.017$ (B3LYP$\corrtemp$).

In (B2), we seek $\theta$ to reproduce the entropy $S$ of the CAS reference at each geometry ($\theta_S$). It is not just a proxy for the occupation number spectrum; several investigations, including our study on H$_4^2+$, posit a direct relationship between correlation energy dominated by static correlation and the entropy of the true ground state. The present approach guarantees that its reference value is recovered.
Since the form of the thermally enhanced functional implies that $S$ and $\theta$ are conjugate variables, the correlation temperature can be interpreted as a Lagrange multiplier that controls the entropy. Hence, $\theta$ can be set such that the reference $S$ is matched exactly for each geometry. 

The resulting sets of $\theta$ are presented in panels b and e of Fig.~\ref{fig:n2_kappa} with purple ($\theta_\mathrm{NON}$) and pink curves ($\theta_\mathrm{S}$).

\subsection{Fits of \texorpdfstring{$\theta$}{theta} to occupation number spectrum}\label{sec:theta-occ-op}

With the third class of methods for computing correlation temperatures, we obtain $\theta$ not by focussing on any outcome of an SCF$\theta$ correlation but, recognizing the central effect of the temperature on thermally enhanced calculations via the Fermi-Dirac (FD) distribution, as the result of a one-shot fit of this distribution to the natural occupation numbers (NONs) from a reference calculation.
This approach is partly motivated by our recent CASSCF study in H$_4^{2+}$, where for specific deformation paths inducing strong static correlation in the ground state, we found the cumulant correlation energy proportional to the particle-hole symmetrized entropy and $\theta$ from an FD fit matching the (negative) slope to a high accuracy.~\cite{perspective}

Besides this, we have two reasons for quantifying the NON patterns via the FD distribution: firstly, it directly results from the form of the correlation functional we propose to use in thermally enhanced methods. This also allows a direct comparison with the values for $\theta$ obtained via methods (A1) \& (A2). Secondly, more fundamentally, the Fermi-Dirac distribution represents the least-biased distribution (in the sense of Jaynes' \emph{MaxEnt} principle~\cite{Jaynes1957}) that is consistent with particle-number and energy constraints. We choose to fit by minimizing the $ell^2$-error of the occupation numbers (similar to what has been done in Section~\ref{sec:theta-occ-op}) rather than by minimizing the binary relative entropy, as MaxEnt would suggest; we also include $\mu$ as a fit parameter. However, additional calculations showed that both of these choices do not influence the resulting values of $\theta$ significantly.

However, before we can apply the FD fit
\begin{equation}\label{eq:FD-distr}
    (\mu_\mathrm{fit},\theta_\mathrm{fit}) = \argmin_{\mu,\theta}\sum_i\left[\left(1+\rme^{(\varepsilon_i-\mu)/\theta}\right)^{-1}-n_i^\mathrm{ref}\right]^2
\end{equation}
to the reference NONs $\{n_i^\mathrm{ref}\}$, we must choose a relevant set of orbital energies $\{\varepsilon_i\}$; note that they do not have a natural definition in the correlated wave function. While the least-squares fit of the orbital occupation numbers is in common with the procedure in Section~\ref{sec:theta-occ-op} (see also Eq.~\eqref{eq:ell2-error-nons}), the orbital energies here are fixed before the fit, and $\mu$ is kept free.
We discuss three distinct ansätze (C1-3) differing in their choice of orbital energies. For brevity, we provide only brief descriptions here and a detailed definition in the Supplementary Material.

In (C1), we use orbital energies computed from the natural orbitals of high-level wave functions analogous to their definition in Ref.~\cite{perspective}, such that the prediction of $\theta$ is carried out purely as an analysis of a correlated ground-state wave function without any mean-field approximation. It is thus also in the spirit of early works on the Collins' conjecture~\cite{Collins,WangKnowles21-PRA,Graf2026} that tried to find an occupation-energy relation based only on high-level calculations. Since the electronic structure of these might differ significantly from that of mean-field calculations whose reliability is the target of our search for $\theta$, we alternatively consider orbital energies from mean-field calculations at zero temperature (C2) or at nonzero temperature (C3). In these cases, we match the increasingly ordered orbital energies with the decreasingly ordered reference occupation numbers. An important decision within method (C3) is the choice of the value for $\theta$, here called $\theta_0$, for the SCF$\corrtemp$ calculations that determine the orbital energies, since it will influence the $\theta_\mathrm{fit}$ determined with Eq.~\eqref{eq:FD-distr}. For our analysis, we choose the values $\theta_S$, obtained via method (B2), which are optimized to reproduce the reference entropy.

Another important choice motivated by previous similar work on occupation number-fitting~\cite{Chen2022} is which occupation numbers to consider. 
Since our positive results in H$_4^{2+}$ are obtained when the occupation spectrum is dominated by few statically correlated orbitals~\cite{perspective}, and the purpose of the fractional occupations is the description of strong static correlation, we restrict our fit to those orbitals experiencing significant changes in occupation throughout the dissociation. In nitrogen, static correlation predominantly affects the 3$\upsigma_g$ and 3$\upsigma_u$ orbitals, as well as the 1$\uppi_u$ and 1$\uppi_g$ (frontier) orbitals. 
For each mean-field functional HF and B3LYP, we carry out 5 calculations belonging to these three categories:
\begin{enumerate}[(C1)]
    \item uses orbital energies derived from the high-level reference calculation; either all six statically correlated orbitals (blue curve), only the frontier 1$\uppi$ orbitals (green curve) or only the 3$\upsigma$ HOMO-2 and LUMO+2 orbitals (orange curve).
    \item uses orbital energies from a reference zero-temperature SCF calculation, where we show only the result of a fit of all six relevant orbital occupations (red curve).
    \item uses orbital energies from SCF$\corrtemp$ calculations, where $\theta_0$ entering the SCF$\corrtemp$ calculation is $\theta_S$, adopted from method (B2). We show only the result of a fit of all six relevant orbital occupations (brown curve).
\end{enumerate}

\subsection{Results in the dissociation of N\texorpdfstring{$_2$}{2}}\label{sec:theta-op-results}

Having defined and discussed the various approaches to obtain $\theta$ in detail, we now compare the results obtained for the dissociation of nitrogen. We recall that the wave function reference used below was obtained via CASSCF(6,24) calculations. The SCF($\corrtemp$) calculations were carried out with HF($\corrtemp$) or i-DMFT (= HF$\corrtemp$ with additional constant term $-b$) in panels a, b, and c, and B3LYP($\corrtemp$) (partly with $-b$) in panels d, e, and f.

First, we turn towards the energy-based optimization schemes for $\theta$ from Section~\ref{sec:theta-en-op} in panels c and f, where both geometry-specific (dashed lines, (A2)) and globally fitted correlation temperatures (solid lines, (A1)) are presented (recall that the global values of $\theta$ are provided in Table~\ref{tab:e_diss_n2_theta_glob}).
As detailed above, the geometry-specific $\theta$ are chosen to match the absolute energy of the CAS reference after inclusion of a constant offset ($-b$) as in the thermal methods. Since the entropy varies significantly during dissociation, the offset greatly affects the stability of the correlation temperature curves.
In fact, only when $b$ is chosen equal to the reference offset, $\Delta E^\mathrm{abs,CAS}$, do we find a low variation in $\theta$ along the dissociation. This substantiates the pragmatic scheme that was used here and in previous works for the global interpolation of $\theta$.
When choosing $b$ differently (or even $b=0$, as in TAO-DFT~\cite{Chai2012}), we always observe divergent behavior towards smaller bond lengths. This holds true both for HF$\corrtemp$ and B3LYP$\corrtemp$. In the latter, a too small offset does not only make $\theta$ vary more along the dissociation coordinate, but also results in a larger range of $R$ where $\theta$ cannot be determined due to the B3LYP$\corrtemp$ energy falling below the reference energy.

Next, we focus on the results for occupation-based definitions of the correlation temperature (Sections~\ref{sec:theta-match-op} \& \ref{sec:theta-occ-op}), which are presented in panels a, b, d and e.
The results in all four panels show similar patterns across most curves, with a minimum around 3 Bohr, followed by an increase and a plateau (panel a) or a gentle slope (panels d and e) from ca.\ 5~Bohr onward. This behavior is matched by all those $\theta$ obtained via approaches (B1) (approximating reference NONs), (B2) (matching reference entropy), (C1) (orbital energies based on CASSCF orbitals) and (C3) (based on SCF$\corrtemp$ orbital energies), with only minor deviations along the dissociation. In the dissociation limit, notably, independent of which subset of active orbitals was used (blue, orange, green), all produce near-identical values for $\theta$ (implying that in the dissociation limit, the statically correlated orbitals all align with a single Fermi-Dirac distribution). A strong deviation is only found in approach (C2), where the Fermi-Dirac fit is applied based on orbital energies from zero-temperature SCF calculations. While yielding reasonable values around the equilibrium geometry, the red curves diverge strongly beyond small bond distances of ca.\ 3~Bohr. This can be explained by the insufficient description of the vanishing HOMO-LUMO energy gap along the dissociation in zero-temperature SCF methods.

The central difference between the results in the four panels is the absolute values of $\theta$, which are significantly lower when using the DFT functional, consistent with the findings of Section~\ref{sec:study-dissoc}.

Compared with the globally fitted values of $\theta$, most of the occupation number fits along the dissociation curve would not reproduce the correct energetics in SCF$\corrtemp$ calculations, even though they might improve the approximation of the occupation pattern.
This underscores the notion that optimizing energy and occupations requires different electronic temperatures and demonstrates that an adiabatic connection hybrid method that preserves orbital occupations in addition to the total density~\cite{han2026} can yield inferior results.
As concerns finding a method for predicting an accurate $\theta$ without explicitly fitting energies, fitting occupation numbers around the experimental equilibrium distance (2.07 Bohr) to mean-field orbital energies allows one to recover the CAS-based global $\theta$ to within about 0.01~Ha. 

In HF$\corrtemp$, the best result was achieved by $\theta$ fixing the correct entropy in SCF$\corrtemp$ ($\theta_S$), where at $R=2.07$~Bohr we obtain $\theta=0.116$~Ha. This is followed by the fit to the HF orbital energies (B2) with $\theta=0.113$~Ha and the fits to HF$\corrtemp$ orbital energies obtained with $\theta_0=\theta_S$ (C3) and $\theta$ approximating occupations (B1), both returning $\theta=0.109$~Ha.
In B3LYP$\corrtemp$, the closest match to the global $\theta$ was obtained from fitting the Fermi-Dirac distribution to NONs over orbital energies from SCF$\corrtemp$ with $\theta_0=\theta_S$ fixing the entropy (C3) or $\theta$ optimizing the occupations (B1), both yielding $0.060$~Ha. Again, the fit based on SCF orbital energies (C2) ($\theta=0.061$~Ha) and the $\theta_S=0.065$~Ha itself (B1) yield reasonable results. All values of $\theta$ at $R=2.07$~Bohr in panels a, b, d and f are given in the Supplementary Material.

The results of these fitting procedures imply two things. First, the assumption of a constant $\theta$ along the dissociation coordinate is only valid if one admits a constant offset of the absolute energy. Secondly, if one is looking for a constant $\theta$ that captures absolute dissociation energetics, it cannot be obtained from a single reference point, since additional knowledge about the offset is needed to make $\theta$ constant. However, in the studied system, we found that the reference points for a global fit could be obtained at the SCF level entirely, by performing single-point B3LYP calculations. 

Furthermore, we confirmed that with the correct procedure, $\theta$ can be fitted at the equilibrium geometry and then be applied along the entire dissociation pathway. Notably, this is also possible based on fits to natural occupation number patterns. Future work may focus on determining these occupation numbers using cheaper wave-function approximations, such as MP2 or CCSD, thereby allowing us to estimate $\theta$ with single-reference methods without explicitly performing CAS or even FCI calculations.

\section{Benchmarks for larger systems}
\label{sec:larger}

These results make it clear that care needs to be taken when choosing the $\theta$ value, where the reproduction of the dissociation energy is a very consistent option. Already in our previous study~\cite{Graf2026}, we hypothesized that $\theta$ is similar for certain bond types due to its correlation with the bond dissociation energy and control mainly over the occupation of orbitals involved in a dissociation. If this is true, $\theta$ can be determined from calculations on chemically similar motifs in smaller molecules and later applied to larger target molecules. In this section, we test this hypothesis.

\begin{figure}
    \centering
    \includegraphics[width=0.21\linewidth]{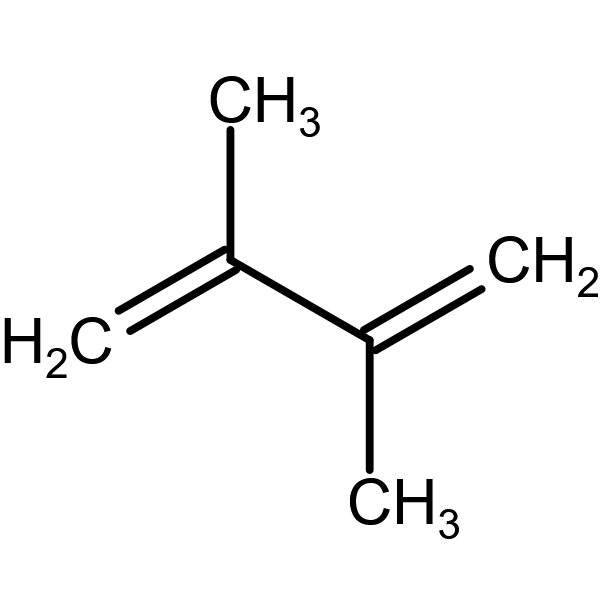}
    \includegraphics[width=0.37\linewidth]{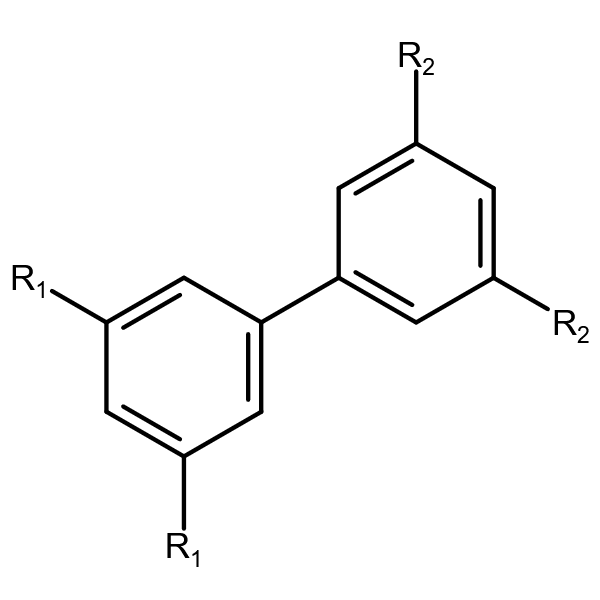}
    \includegraphics[width=0.39\linewidth]{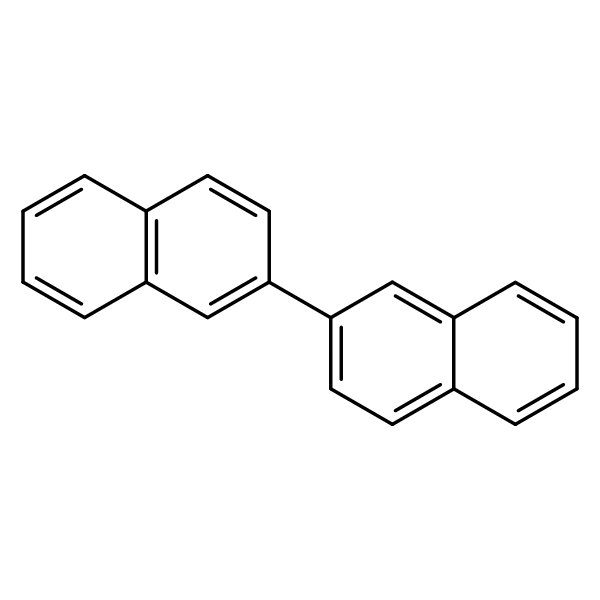}
    \caption{Investigated molecules \textbf{1} (left), \textbf{2}-\textbf{8} (middle), and \textbf{9} (right).}
    \label{fig:structures}
\end{figure}

We present the results of a benchmark of the method on nine different molecules, in which we dissociate single bonds between sp$^2$ hybridized carbon atoms. Their bond strengths are particularly sensitive to a description of electron correlation associated with $\uppi$--$\uppi$ conjugation, which is prevalent in many relevant organic molecules. In particular, we chose one small nonaromatic example (2,3-dimethylbutadiene, \textbf{1}), seven differently substituted biphenyls (\textbf{2}-\textbf{8}), and 2-2'-binaphthyl (\textbf{9}). The structures of the test systems are shown in Fig.~\ref{fig:structures}, and the different residues are listed in Table~\ref{tab:residues}.

\begin{table}
    \centering
    \caption{Substitution patterns in biphenyl molecules \textbf{2}-\textbf{8}.}
    \begin{tabular}{lrr}\toprule
        Molecule&R$_1$&R$_2$\\\midrule
        \textbf{2}&H&H\\
        \textbf{3}&F&F\\
        \textbf{4}&CH$_3$&CH$_3$\\
        \textbf{5}&OH&OH\\
        \textbf{6}&CHO&CHO\\
        \textbf{7}&CH$_3$&CHO\\
        \textbf{8}&F&OH\\\bottomrule
    \end{tabular}
    \label{tab:residues}
\end{table}

The geometries used for calculations on these molecules are available from the Supplementary Material. We used the PySCF software package to carry out CASSCF calculations with 6 active electrons in 6 active orbitals for molecule \textbf{1} and 14 active electrons in 14 active orbitals for the other molecules as well as CCSD calculations, each using the cc-pVDZ basis set. The CASSCF calculations serve to establish a good reference value for the dissociation energy, while the CCSD calculations are a second reference that offers a correct description of dynamic correlation around the equilibrium distance.

In every case, we computed energies for 80 geometries representing a symmetric dissociation of the molecule which only differ in the bond distance between the two central carbon atoms, which was sampled from 0.28 to 9.45 Bohr in steps of 0.116 Bohr. In molecule \textbf{1}, the four carbon atoms are all coplanar, while the aromatic rings are twisted against each other by ca. 40$^\circ$ in molecules \textbf{2}-\textbf{9}, such that the symmetrically substituted molecules have D$_2$ point group symmetry. The dissociation yields two radical systems, which we treat as singlet-coupled: the compound system stays a singlet along the dissociation. 

\begin{figure}
    \centering
    \includegraphics[width=\linewidth]{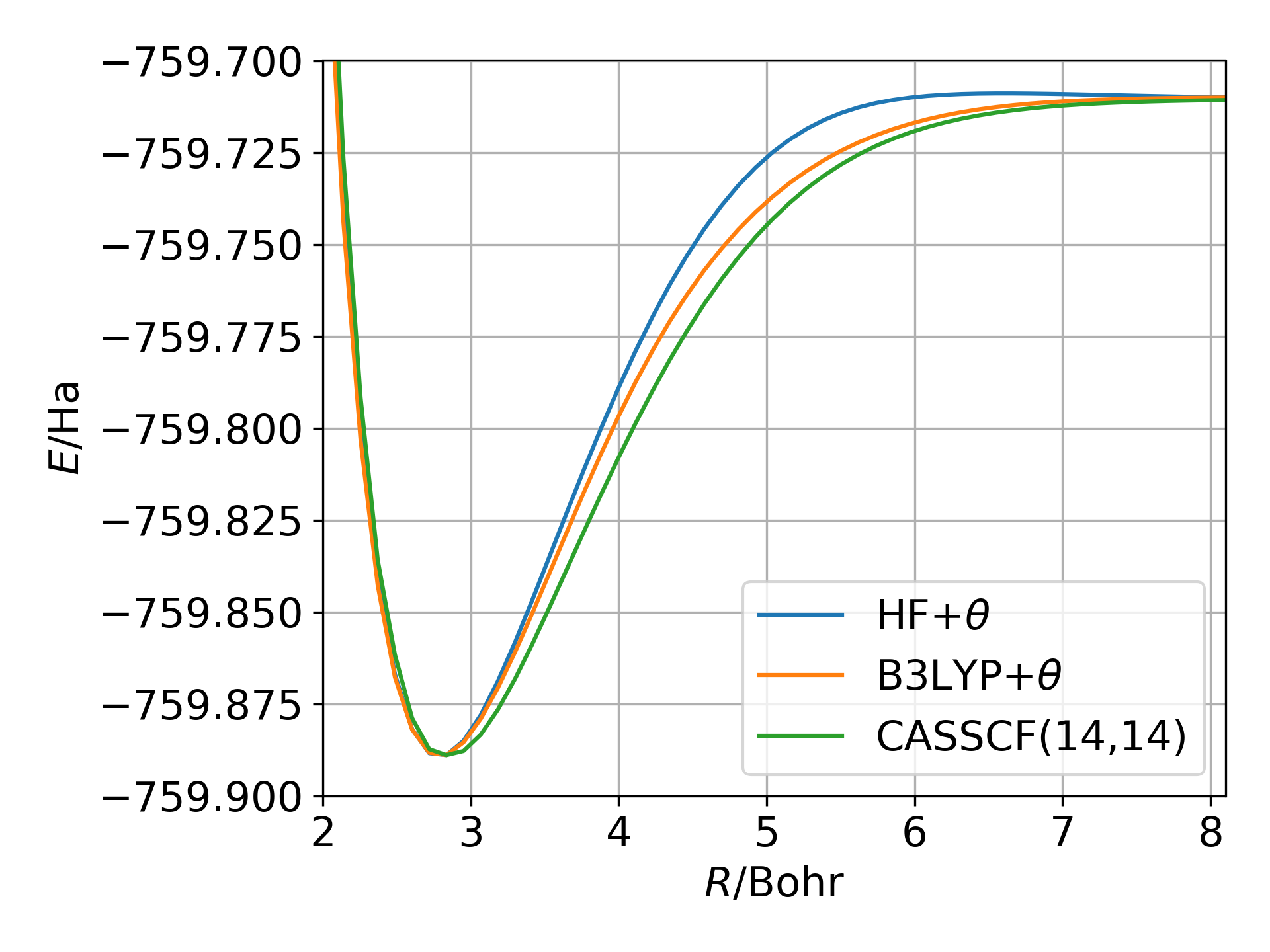}
    \caption{Dissociation curve for molecule \textbf{5}, computed with CASSCF(14,14), HF$\corrtemp$, and B3LYP$\corrtemp$ (the latter two methods with optimized $\theta$), in each case in the cc-pVDZ basis set. Curves shifted to align the energy minima.}
    \label{fig:bp-ohoh-e}
\end{figure}

\begin{figure}
    \centering
    \includegraphics[width=\linewidth]{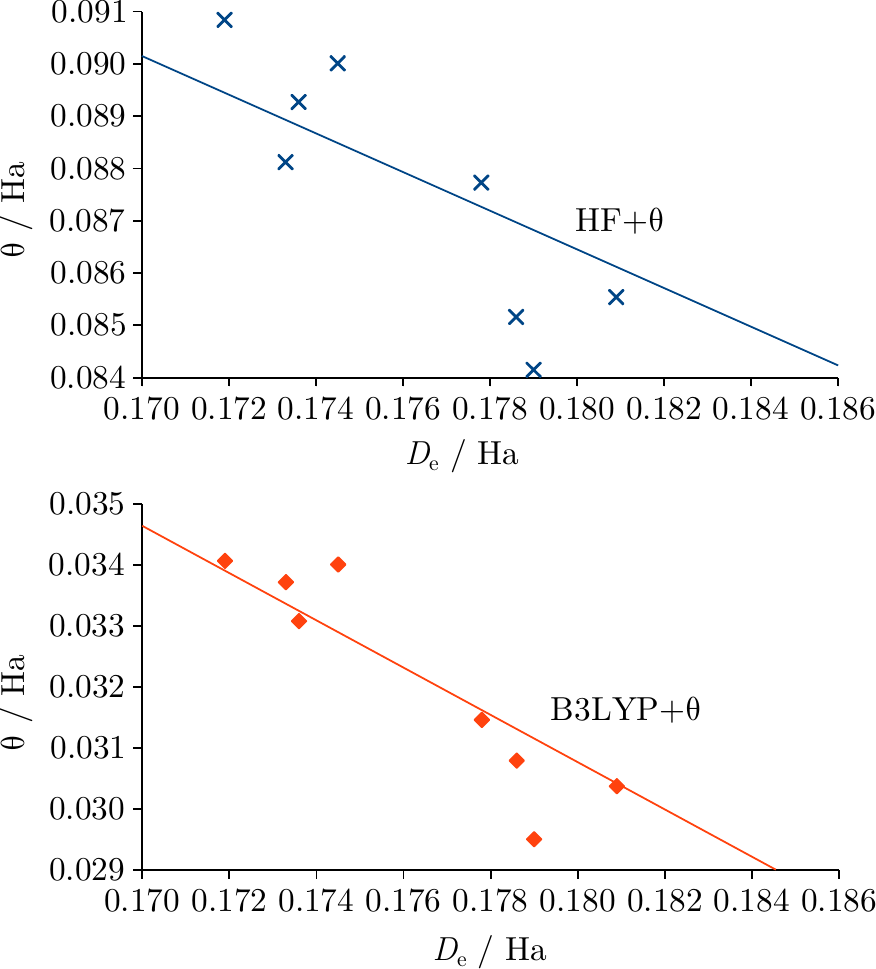}
    \caption{Optimized $\theta$ for molecules \textbf{1}-\textbf{9} plotted against the respective dissociation energies.}
    \label{fig:kappas}
\end{figure}

From the computed dissociation curves, we determined the dissociation energy $D_\rme$, the optimal bond distance $r_\rme$, and the bond stiffness parameter $w$ as per the fitting procedure outlined in the Supplementary Material. This data is listed in Table~\ref{tab:biphenyls-data}. Additionally, we carried out both HF$\corrtemp$ and B3LYP$\corrtemp$ calculations, where for each functional and system, $\theta$ was chosen to match the CASSCF dissociation energy. An example for dissociation curves obtained with different methods is shown for molecule \textbf{5} in Fig.~\ref{fig:bp-ohoh-e}. Table~\ref{tab:biphenyls-data} shows the optimized $\theta$ values as well as the other quantities computed at the optimal value of $\theta$. The values obtained for $\theta$ in comparison to the dissociation energies are also shown in Fig.~\ref{fig:kappas}.

\begin{table*}
    \centering
    \caption{Results of benchmark calculations with different methods; $\theta$ in the thermally enhanced methods was chosen to reproduce the CASSCF dissociation energy. $D_\rme$ and $\theta$ in $\mathrm{Ha}$, $r_\rme$ in Bohr, $w$ in Bohr$^{-2}$.}
    \begin{tabular}{lrrrrrrrrrrr}\toprule
        System & \multicolumn{3}{c}{CASSCF} & \multicolumn{2}{c}{CCSD} & \multicolumn{3}{c}{HF$\corrtemp$} & \multicolumn{3}{c}{B3LYP$\corrtemp$}\\
        &$D_\rme$ & $r_\rme$ & $w$ & $r_\rme$ & $w$ & $\theta$ & $r_\rme$ & $w$ & $\theta$ & $r_\rme$ & $w$\\\midrule
        \textbf{1}&0.1719&2.8338&0.9295&2.8296&0.8819&0.09084&2.8044&0.9508&0.03407&2.7980&0.9471\\
        \textbf{2}&0.1745&2.8703&0.8330&2.8378&0.8757&0.09001&2.7926&0.9383&0.03401&2.7977&0.9351\\
        \textbf{3}&0.1733&2.8914&0.6948&2.8258&0.8725&0.08812&2.7832&0.9378&0.03372&2.7851&0.9432\\
        \textbf{4}&0.1736&2.8525&0.9113&2.8372&0.8844&0.08927&2.7945&0.9365&0.03308&2.7992&0.9345\\
        \textbf{5}&0.1790&2.8481&0.8685&2.8276&0.8853&0.08415&2.7886&0.9242&0.02950&2.7933&0.9304\\
        \textbf{6}&0.1866&2.8060&1.0081&2.8288&0.8868&0.08605&2.8018&0.9134&0.02912&2.8105&0.9114\\
        \textbf{7}&0.1778&2.8306&1.0488&2.8362&0.8847&0.08773&2.7953&0.9305&0.03146&2.8016&0.9285\\
        \textbf{8}&0.1786&2.8557&0.8884&2.8267&0.8734&0.08516&2.7859&0.9277&0.03079&2.7900&0.9337\\
        \textbf{9}&0.1809&2.8123&1.0172&2.8270&0.8827&0.08554&2.7973&0.9151&0.03037&2.8037&0.9156\\
        \bottomrule
    \end{tabular}
    \label{tab:biphenyls-data}
\end{table*}

The optimized values for $\theta$ vary slightly from molecule to molecule. Similar to before, they are much lower for the B3LYP functional than for the HF functional. The average values for $\theta$ are 0.0870 $\mathrm{Ha}$ with HF and 0.0315 $\mathrm{Ha}$ with B3LYP, and the standard deviations are 0.0021 $\mathrm{Ha}$ (2.4\%) and 0.0019 $\mathrm{Ha}$ (6.0\%), respectively. In general, $\theta$ follows a similar trend when comparing results for the two methods, e.g., the nonaromatic butadiene \textbf{1} requiring the largest $\theta$, coinciding with the lowest bond strength. In general, such a connection between bond strength and $\theta$ is apparent as illustrated in Fig.~\ref{fig:kappas} even though the correlation is not perfect. The slope of the least-squares linear fit is $-0.370$ for HF$\corrtemp$ and $-0.388$ with B3LYP$\corrtemp$, notably similar to the value $-1/(4\log 2) \approx -0.361$, where $4\log 2 = -4\cdot(\frac{1}{2}\log\frac{1}{2}+(1-\frac{1}{2})\log(1-\frac{1}{2}))$ is the entropy gain associated with a single bond breaking (transition from orbital occupations $1,1,0,0$ to $\frac{1}{2},\frac{1}{2},\frac{1}{2},\frac{1}{2}$). This implies that the mean-field energies on their own do not model the difference in bond energies between the molecules, such that it needs to be corrected by introducing different $\theta$ values, leading to an approximately proportional energy correction for the dissociated molecule.

The potential energy surface shown in Fig.~\ref{fig:bp-ohoh-e} shows a similar behavior as the one for ethylene discussed in Section~\ref{sec:study-dissoc:ethylene}: the HF$\corrtemp$ calculation exhibits a small artificial dissociation barrier, which is corrected by B3LYP$\corrtemp$. However, a quantitative analysis of the fitting parameters shows that both thermally enhanced methods underestimate the optimal bond distances compared to CASSCF and CCSD results and do not even reproduce the trend between the molecules. The bond stiffness parameters $w$ are larger than those obtained with CCSD, while the CASSCF results vary strongly from molecule to molecule, potentially an artifact of a switch between different CASSCF roots. For both quantities, using a correlation DFA only leads to a minimal improvement over simply using the HF functional.

Finally, we repeated the thermally enhanced calculations on the 9 test systems with the same value of $\theta$, that was determined as the average of the $\theta$ values optimized for matching the dissociation energies. This is in the spirit of many previous works on TAO-DFT which tried to find general values for $\theta$ or at least values that hold for multiple molecules.~\cite{Chai2012,Chai2014,Chai2017,Lin2017,Chen2022} These results are shown in Table~\ref{tab:biphenyls-dft-results}.

\begin{table*}
    \centering
    \caption{Results of benchmark calculations with different methods; $\theta$ in all thermally enhanced methods was chosen as the average value of the previously optimized values. $D_\rme$ and $\theta$ in $\mathrm{Ha}$, $r_\rme$ in Bohr, $w$ in Bohr$^{-2}$.}
    \begin{tabular}{lrrrrrrrrr}\toprule
        System & \multicolumn{3}{c}{CASSCF} & \multicolumn{3}{c}{HF$\corrtemp$} & \multicolumn{3}{c}{B3LYP$\corrtemp$}\\
        & $D_\rme$ & $r_\rme$ & $w$ & $D_\rme$ & $r_\rme$ & $w$ & $D_\rme$ & $r_\rme$ & $w$\\\midrule
        \textbf{1}&0.1719&2.8338&0.9295&0.1805&2.8048&0.9390&0.1794&2.8010&0.9382\\
        \textbf{2}&0.1745&2.8703&0.8330&0.1810&2.7935&0.9291&0.1817&2.8017&0.9264\\
        \textbf{3}&0.1733&2.8914&0.6948&0.1751&2.7834&0.9352&0.1801&2.7888&0.9335\\
        \textbf{4}&0.1736&2.8525&0.9113&0.1782&2.7951&0.9300&0.1781&2.8017&0.9291\\
        \textbf{5}&0.1790&2.8481&0.8685&0.1710&2.7873&0.9359&0.1729&2.7900&0.9382\\
        \textbf{6}&0.1866&2.8060&1.0081&0.1831&2.8015&0.9179&0.1801&2.8079&0.9173\\
        \textbf{7}&0.1778&2.8306&1.0488&0.1785&2.7954&0.9295&0.1777&2.8015&0.9286\\
        \textbf{8}&0.1786&2.8557&0.8884&0.1729&2.7851&0.9360&0.1764&2.7888&0.9366\\
        \textbf{9}&0.1809&2.8123&1.0172&0.1764&2.7969&0.9214&0.1773&2.8025&0.9190\\\bottomrule
    \end{tabular}
    \label{tab:biphenyls-dft-results}
\end{table*}

With a uniform $\theta$, the methods predict the dissociation energies with an accuracy of about 0.01 $\mathrm{Ha}$, but cannot establish the correct order of bond strengths among the different molecules as expected based on the discussion of Fig.~\ref{fig:kappas}. Beyond the energy, the thermally enhanced methods still fail to predict the variations in bond stiffness and equilibrium distance between the molecules, which seems to be a more general problem of the thermally enhanced methods.

\section{Conclusion}
\label{sec:conclusions}

In this manuscript, we showed how thermally enhanced hybrid KS density functionals can be constructed via the thermally enhanced adiabatic connection and applied to a variety of quantum-chemical problems, where they yield an accurate description of static and dynamic correlation at mean-field cost. 

While DFT methods with correlation temperature have been proposed and tested before (see citations in Section~\ref{sec:intro}), the methodology presented here goes beyond using electronic temperature merely as a numerical technique to improve convergence, and instead derives an entropic contribution to the functional approximation from first principles. 
Since the entropy is a function of the reduced one-particle density matrix, this allows us to establish a clear connection to quantum information theory and reduced density matrix functional theory.~\cite{perspective} The motivation behind the development of this variant of DFT is that state-of-the-art KS-DFT is well capable of describing dynamic correlation within a single-reference picture, but cannot describe multiconfigurational wave functions. Thermal electronic structure methods, on the other hand, have a proxy (the entropy) that describes static correlation among orbitals with fractional occupations, while dynamic correlation remains too complicated to be reduced to a single scalar. 

Our derivation started from well-established DFT principles. Hybrid functionals combine HF (``exact'') exchange with density-based approximations that already do well at managing dynamic correlation. The adiabatic connection formalism that enables this form of approximation then delivers thermally enhanced hybrid functionals, which can possibly be improved in several manners. The combination with an LDA functional at finite temperature~\cite{Dornheim2018} would add an additional anchor point in the thermally enhanced adiabatic connection. Similarly, the explicit consideration of the strongly interacting limit~\cite{GoriGiorgi2009} can allow for a further expansion of the theory. Another possible direction is the extension of the adiabatic-connection fluctuation-dissipation theorem to temperature.

As our numerical results show, the new hybrid method can reproduce electron density, dissociation curves, and natural occupation numbers much more faithfully than the previously proposed i-DMFT method based only on the HF functional. The error in the computed absolute energies is also reduced by factors between 3 and 300. In our case, the calculations were carried out using established parametrized density functionals; it is likely that the quality of the results can further be improved when reparameterizing the exchange-correlation functionals by adapting the weight $a$ to the effective temperature $\theta$. We also investigated a range-separated hybrid functional based on the $\omega$B97X-V DFA, which performed similarly well for hydrogen and ethylene, but showed larger deviations in nitrogen. While range-separated functionals are highly popular in modern DFT, they might require a specific treatment when combining with entropy-based functionals, as for example by means of a range-separated evaluation of the entropy.

It needs to be emphasized that the choice of an appropriate correlation temperature is still the main challenge for the application of thermally enhanced methods. This is most relevant in the presence of strong correlation, where the temperature contributes significantly to the energy functional. This motivated our detailed investigation of procedures for an ab initio determination of $\theta$ in Section~\ref{sec:theta-approaches} that also discussed approaches where it needs to be varied along a dissociation curve, calling the assumption of constant or universal values of $\theta$ into question. From these results emerge two main recipes that can yield quantitatively correct dissociation curves. The fitting of the dissociation energy to a reference calculation, which can also be a low-level DFT calculation (in which case the thermally enhanced method serves to interpolate the dissociation curve, which is achieved at true mean-field cost); and the fitting of natural occupation numbers to the results of a high-level method, where already the relatively simplistic electronic structure of a molecule in its equilibrium structure can enable predictions about its dissociation energetics.

The benchmarking on a series of organic molecules in Section~\ref{sec:larger} revealed a fundamental challenge: even for chemically similar dissociating bonds, $\theta$ cannot be assumed constant if relative dissociation energies matter. On the other hand, it should be noted that the description can be significantly improved with respect to a zero-temperature single-determinant KS or HF calculation even with an approximate $\theta$ applied to distinct but similar molecules. 

Overall, we see considerable potential in thermally enhanced density-functional methods for quantum chemistry. We hope that the data, tools, and strategies presented here will pave the way for applying these methods to more challenging problems. More generally, we conclude that a rigorous theoretical foundation not only ensures the validity of computations but also points to systematic ways of improving quantum-chemical methods.

\begin{acknowledgments}
We thank Vebj\o{}rn H.\ Bakkestuen and Pierre-Olivier Roy for helpful comments on the manuscript. We acknowledge financial support from the German Research Foundation (DFG, Grant No.~SCHI~1476/1-1) (P.G.G., F.M., M.P., C.S.) and from the European Research Council (ERC) under the Starting Grant REGAL (ERC-2021-STG, grant agreement No.~101041487) (M.P.) and the Consolidator Grant beyondDFT (ERC-2025-COG, grant agreement No.~101231616) (C.S.). This work is part of the Munich Quantum Valley, which is supported by the Bavarian state government with funds from the Hightech Agenda Bayern Plus.
\end{acknowledgments}

\medskip
\section*{Statement on the use of generative AI}
The authors used Claude Opus 5.5 (Anthropic) to improve the language and arguments in the manuscript through iterative revisions based on the authors' content and feedback. All AI-generated text was reviewed and verified by the authors, who take full responsibility for the content.

\bibliography{main}

\end{document}